%% file: main.tex
\documentclass[%
 reprint,
 amsmath,amssymb,
 aps,
 prx,
]{revtex4-2}

\input{packages.tex}

\newcommandx{\figref}[2][2=emptyarg]{\ifthenelse{\equal{#2}{emptyarg}}{Fig.~\ref{#1}}{Fig.~\hyperref[#1]{\ref*{#1}\textcolor{blue}{({#2})}}}}

\begin{document}

\title{Edge modes and symmetry-protected topological states in open quantum systems}
\author{Dawid~Paszko}
\author{Dominic~C.~Rose}
\author{Marzena~H.~Szymańska}
\author{Arijeet~Pal}
\affiliation{Department of Physics and Astronomy, University College London, Gower Street, London WC1E 6BT, United Kingdom}

\begin{abstract}
\input{0_abstract}

\end{abstract}

\maketitle

\section{Introduction}
\label{sec:intro}
\input{1_introduction}


\section{Framework}
\input{2_model}

\section{Strong symmetries in the master equation} \label{sec:ss}
\input{3_strong}

\section{Weak symmetries in trajectories}\label{sec:traj}
\input{4_weak}


\section{Conclusion and outlook}\label{sec:conclusion}
\input{9_conclusion}

\appendix

\input{A1_mathdetails}




\bibliography{Bibliography}

\end{document}

%% file: packages.tex
\usepackage[utf8]{inputenc} 
\usepackage{graphicx}

\usepackage{csquotes}

\usepackage{fancyhdr}
\usepackage{float}
\usepackage{amssymb}
\usepackage{upgreek}
\usepackage{amsmath}
\usepackage{indentfirst}

\usepackage{dsfont}

\usepackage[english]{babel}
\usepackage[usenames, dvipsnames]{color}
\usepackage{soul}
\usepackage{multirow}
\usepackage{ragged2e}
\usepackage{mdframed}
\usepackage{gensymb}
\usepackage{physics}
\usepackage{comment}
\usepackage{emptypage}

\usepackage{enumitem}

\usepackage{pgfplots}
\usepgfplotslibrary{fillbetween,groupplots,external}
\pgfplotsset{compat=newest}
\pgfplotsset{plot coordinates/math parser=false}
\newlength\figureheight
\newlength\figurewidth 

\pgfplotsset{aspect ratio/.code args={#1:#2}{%
		\def\axisdefaultwidth{#1 cm}%
		\def\axisdefaultheight{#2 cm}},
	compat=newest
}

\pgfplotsset{
        layers/my layer set/.define layer set={
            background,
            main,
            foreground
        }{
        },
        set layers=my layer set,
    }
    
\usepackage{bm}

\usepackage[colorlinks = true, citecolor = blue, linkcolor = blue, urlcolor=blue]{hyperref}

\usepackage{xcolor}

\usepackage{tikz}
\usepackage{tikz-3dplot}
\usetikzlibrary{3d,spy,external,matrix,plotmarks,pgfplots.groupplots,calc,decorations.pathreplacing, decorations.pathmorphing, arrows, decorations.markings, quantikz, fadings}

\newmdenv[outerlinewidth=1,outerlinecolor=grey,frametitle=Démonstration,innertopmargin=10pt,leftmargin=0,rightmargin=0]{proof}

\newmdenv[outerlinewidth=1,outerlinecolor=grey,innertopmargin=10pt,leftmargin=0,rightmargin=0]{customtitle}

\newmdenv[linewidth=1,linecolor=grey,innertopmargin=10pt,leftmargin=40,rightmargin=40]{see}

\let\oldhat\hat
\renewcommand{\vec}[1]{\bm{\mathbf{#1}}}
\renewcommand{\hat}[1]{\oldhat{\bm{\mathbf{#1}}}}

\usepackage[T1]{fontenc} 

\pgfdeclaredecoration{complete sines}{initial}
{
	\state{initial}[
	width=+0pt,
	next state=sine,
	persistent precomputation={\pgfmathsetmacro\matchinglength{
			\pgfdecoratedinputsegmentlength / int(\pgfdecoratedinputsegmentlength/\pgfdecorationsegmentlength)}
		\setlength{\pgfdecorationsegmentlength}{\matchinglength pt}
	}] {}
	\state{sine}[width=\pgfdecorationsegmentlength]{
		\pgfpathsine{\pgfpoint{0.25\pgfdecorationsegmentlength}{0.5\pgfdecorationsegmentamplitude}}
		\pgfpathcosine{\pgfpoint{0.25\pgfdecorationsegmentlength}{-0.5\pgfdecorationsegmentamplitude}}
		\pgfpathsine{\pgfpoint{0.25\pgfdecorationsegmentlength}{-0.5\pgfdecorationsegmentamplitude}}
		\pgfpathcosine{\pgfpoint{0.25\pgfdecorationsegmentlength}{0.5\pgfdecorationsegmentamplitude}}
	}
	\state{final}{}
}

\tikzset{
	photon/.style={decorate, decoration={
			complete sines,
			segment length=2.5mm,
			amplitude=1.5mm
		}, draw=black},
	particle/.style={draw=black, postaction={decorate},
		decoration={markings,mark=at position .5 with {\arrow[draw=black]{>}}}},
	antiparticle/.style={draw=black, postaction={decorate},
		decoration={markings,mark=at position .5 with {\arrow[draw=black]{<}}}},
}

\usepackage{graphicx}
\usepackage{dcolumn}

\usepackage{textcomp}
\usepackage{eurosym}
\def\plusheight{-\the\dimexpr\fontdimen22\textfont2\relax}

\pgfplotsset{
compat=1.11,
legend image code/.code={
\draw[mark repeat=2,mark phase=2]
plot coordinates {
(0cm,0cm)
(0.15cm,0cm)        
(0.3cm,0cm)         
};%
}
}

\usepackage{blochsphere}

\pgfplotsset{cell picture=if necessary}

\usepackage{xargs}
\usepackage{ifthen}

\usepackage{tikz-cd}

%% file: 0_abstract.tex
Topological order offers possibilities for processing quantum information which can be immune to imperfections. 
However, the question of its stability out of equilibrium is relevant for experiments, where coupling to an environment is unavoidable.
In this work we demonstrate the robustness of certain aspects of $Z_2 \times Z_2$ symmetry-protected topological (SPT) order against a wide class of dissipation channels in the Lindblad and quantum trajectory formalisms of an open quantum system. 
This is illustrated using the one-dimensional $ZXZ$ cluster Hamiltonian along with Pauli-string jump operators.
We show that certain choices of dissipation retaining strong symmetries support a steady-state manifold consisting of two non-local \textit{logical} qubits, and for Hamiltonian perturbations preserving the global symmetry, states in this manifold remain metastable. 
In contrast, this metastability is destroyed upon breaking the above-mentioned symmetry.
While the localized \textit{edge} qubits of the cluster Hamiltonian are not conserved by the Lindbladian evolution, they do correspond to weak symmetries and thus retain a memory of their initial state at all times in the quantum trajectories.
We utilize this feature to construct protocols to retrieve the quantum information either by monitoring jumps or error mitigation.
Our work thus proposes a novel framework to study the dynamics of dissipative SPT phases and opens the possibility of engineering entangled states relevant to quantum information processing.

%% file: 1_introduction.tex
Dissipation can disrupt entanglement in many-body quantum systems and destroy the physical manifestations of entangled states~\cite{Zurek_RMP2003, Dur_Briegel_PRL2004, barreiro2010_multipartite}. Understanding how to reduce the sensitivity of entangled states to noise is a central challenge for building the next generation of quantum sensors and quantum computers. A unique avenue for protecting entanglement is many-body topological states, encoding the information in non-local motifs of the state and providing robustness to local perturbations~\cite{RevModPhys.80.1083,PhysRevLett.86.5188,PhysRevLett.105.040501,PhysRevLett.119.010504,PhysRevLett.125.120502,PhysRevApplied.17.024057}. However, the extraordinary properties of such topological materials must survive in imperfect conditions, in the presence of disorder ~\cite{Huse_2013_MBLOrder,Bahri2015,PhysRevLett.125.200506} and dissipative coupling to an environment~\cite{Verstraete2009, Diehl2011, Bardyn_2013, PhysRevA.78.042307, mcginley2020fragility}, to be useful in realizing a reliable quantum computer. Indeed, progress towards fault-tolerant computation could potentially be accelerated by the discovery of topologically protected qubits in more realistic systems.

Persistence of topological order -- such as edge modes which act as qubits~\cite{sarma2015majorana, Aasen_2016_Majorana} -- has been observed to be stable and, in fact, may be strengthened in the presence of disorder in prototypical topological many-body systems such as the Kitaev Majorana and Haldane spin-$1$ chains~\cite{parameswaran2018_MBLSymTop}.
In such isolated systems, the entanglement structure of the bulk degrees of freedom provides a fabric which mediates the stability of the edge qubits~\cite{Gu2009_SPT_entang, PhysRevB.81.064439}.
In contrast, the presence of a dissipative environment disrupts the entanglement which supports the edge modes, leading to the loss of coherence.
Dissipation appears to pose severe issues for the physical realization of topological order, which requires a significant level of isolation, giving rise to experimental challenges~\cite{DeLeseleuc2019, Sompet2022, zhang2022digitalFTop,mi2022Edgemodes, jack2019Edgemodes}.

This conflict with dissipation has spurred recent interest in defining indicators of topological order in mixed states which are driven out-of-equilibrium~\cite{PhysRevB.91.165140,PhysRevX.8.011035,bao2023mixedstate,PhysRevResearch.5.023004,deGroot2022symmetryprotected} and attempts have been made to classify symmetry-protected topological (SPT) phases~\cite{PhysRevLett.124.040401,kawabata2022symmetry,sá2022symmetry, lee2022symmetry,zhang2022strange,verissimo2022dissipative}. 
Moreover, engineered forms of dissipation~\cite{Verstraete2009,Diehl2011,Bardyn_2013,Harrington2022,yang2023dissipative} or measurements~\cite{Lavasani2021,morralyepes2023detecting, angelidi2023SPT} have been proposed as a tool to prepare desired topological steady states. 
The detection and behaviour of SPT phases in the dynamics of open system may provide novel mechanisms for preparing and preserving entangled states, necessary for quantum information processing.

Symmetries play an important role in the protection of topological order in isolated systems, providing a route to classifying states of matter in and out of equilibrium~\cite{Chen_2011, Pollmann_2012, Senthil_2015, McGinley_PRL2018, harper2020FloquetTop, PhysRevLett.125.200506, PhysRevB.104.014424}.
In open quantum systems modelled by the Lindblad master equations~\cite{Gorini1976May,Lindblad1976}, symmetries can be either \textit{weak} or \textit{strong} depending on the interplay between the coherent Hamiltonian terms and the dissipative jump operators, leading to rich behaviour~\cite{PhysRevA.89.022118}.
Here, we will analyse how signatures of topological order can persist through these distinct types of symmetries.
 
For concreteness we study the Haldane phase of the cluster model, a well-studied example of an SPT phase, which involves a one-dimensional spin-half chain coupled by three-body interactions~\cite{PhysRevLett.59.799,PhysRevB.45.304,Kennedy1992}. 
This model is relevant to condensed matter systems in magnetism and topological superconductivity and also closely connected to measurement-based quantum computation (MBQC)~\cite{PhysRevA.68.022312,Briegel2009Jan,PhysRevLett.103.020506,PhysRevA.80.022316}. 
We characterize the topology of states using a range of diagnostics including the string order parameter, edge modes, and entanglement spectrum.

Our findings can be summarized as follows.
Firstly, we focus on the effects of strong symmetries at the master equation level.
We illustrate the formation of a non-trivial, degenerate steady-state subspace using dissipators that preserve a strong $\mathds{Z}_2\cross\mathds{Z}_2$ symmetry. 
The stationary state manifold is shown to possess non-trivial string order, along with entanglement spectrum degeneracy typical of topological states.
Further, we show it can act as a quantum memory by encoding a pair of qubits in a non-local manner, which remain despite the presence of dissipation.
We then demonstrate that these qubits remain long-lived in the thermodynamic limit even when affected by perturbations, provided they respect the $\mathds{Z}_2\cross\mathds{Z}_2$ symmetry: we refer to this as symmetry-protected metastability.

Secondly, we consider the effects of weak symmetries through the quantum trajectories formalism~\cite{Daley2014,Macieszczak2021a}, revealing a richer class of topological dynamics invisible to the trajectory-averaged Lindbladian dynamics.
We prove that not only strong symmetries, but also weak symmetries have a striking impact on the stability of topological order. 
We demonstrate that the SPT ordered cluster states, when subjected to such weakly symmetric dissipators, preserve their topological character along trajectory dynamics.
Significantly, we show that the quantum information contained in edge mode qubits is preserved within individual trajectories.
Moreover, we show that this information can be recovered by utilizing the weak symmetry properties, for example, by monitoring quantum jumps happening in only a very small part of the system.

In contrast to previous work which characterised markers of topological order in open systems~\cite{PhysRevB.91.165140,PhysRevX.8.011035,bao2023mixedstate,PhysRevResearch.5.023004,deGroot2022symmetryprotected}, we have gone beyond the mixed-state description by investigating the quantum trajectories of pure states. Moreover, while Ref.~\cite{deGroot2022symmetryprotected} shows that strong symmetries allow string order to survive for a finite time in mixed states -- as opposed to decaying after a single discrete time step for dissipation characterized by weak symmetries  -- we provide a class of models where the string order parameter in a conserved quantity. 
We proceed to study the resulting non-trivial steady state, in particular exploring its potential application as a quantum memory. 
Building on this, we find analogous phenomena in quantum trajectories under the weaker requirement that the dissipation satisfies weak symmetries.
Our results provide a physical mechanism for hosting stationary qubits, even in the presence of dissipation, as a consequence of weak or strong symmetries.

%% file: 2_model.tex
Prior work has focused on preparing - either by engineered dissipation~\cite{Verstraete2009,Diehl2011,Bardyn_2013,Harrington2022,yang2023dissipative} or measurements~\cite{Lavasani2021,morralyepes2023detecting} - topologically ordered steady states, or characterizing this order in mixed steady states~\cite{lee2022symmetry,verissimo2022dissipative}.
In contrast, here we will also focus on demonstrating the dynamical behaviour of SPT phases under dissipation with various types of symmetry, and their robustness to dissipative dynamics. 

To this end, we consider two related frameworks: the Gorini–Kossakowski–Sudarshan–Lindblad master equation~\cite{Gorini1976May,Lindblad1976} (referred to as the Lindbladian in the following), the spectral properties of which will be used to describe stationary mixed states and long-time dynamics; and quantum trajectories~\cite{PhysRevA.46.4382,PhysRevLett.68.580,Molmer93,RevModPhys.70.101,Daley2014}, which provide access to dynamical statistical behavior invisible in the Lindbladian.
These complementary pictures will provide two lenses through which to highlight the various ways quantum information behaves in the models considered.

The Lindblad master equation for the evolution of the density matrix representing the state of the open quantum system is (with $\hbar$ set to $1$ throughout)
\begin{equation}
	\frac{d\rho}{dt}=\mathcal{L}(\rho)= \mathcal{U}(\rho)+ \mathcal{D}(\rho),
	\label{eq:lindblad}
\end{equation}
and consists of a unitary part governed by a Hamiltonian $H$: $\mathcal{U}(\rho)=-i\left[H,\rho\right]$, as well as of a dissipative part modelling the coupling to an environment through a set of jump operators $F_l$: $\mathcal{D}(\rho) = \kappa\sum_{l}\left(2F_l\rho F_l^\dagger -\{F_l^\dagger F_l, \rho \} \right)$.

Identical dynamics can be recovered by averaging over a set of stochastic quantum trajectories. The procedure to generate an individual trajectory consists of (a) evolving an initial pure state $\ket{\psi_0}$ according to the Schroedinger equation with the non-Hermitian effective Hamiltonian $H_{\text{eff}}=H-i\kappa \sum_l F_l^\dagger F_l$, (b) performing randomly chosen quantum jumps at random times $\ket{\psi_t} \longrightarrow F_l\ket{\psi_t}/||F_l\ket{\psi_t}||$, and (c) repeating these two steps in an alternating manner. 
The advantages of this approach are that the state remains pure along a single trajectory and, additionally, that it gives an intuitive physical interpretation about the effect of the environment on the actual dynamics of the state~\cite{Daley2014}. If the environment were monitored such that we could recognize jumps (being thus somehow observable in the environment) occurring in the system, we could assign to it a pure state at all times of the evolution. If not, it is a mixed state (\textit{i.e.} stochastic average over trajectories) that describes our knowledge of the system.

These two perspectives have non-trivial properties in the presence of symmetries, and contrary to closed systems, the definition of symmetry is not straightforward and is not identically related to conserved quantities~\cite{PhysRevA.89.022118}. The strongest condition is that an operator $J$ commutes with both $H$ and $F_l$: $\left[H,J\right]=0, \; \left[F_l,J\right]=0 \; \forall l$. This is called a strong symmetry. It implies two things. First, the operator $J$ is a conserved quantity, $\dot{J}=\mathcal{L}^\dagger(J)=0$, where $\mathcal{L}^\dagger$ determines the evolution of the operator in the Heisenberg picture. The reason is $J$ commutes with everything in $\mathcal{L}^\dagger$. Second, $J$ also generates a symmetry $U=e^{i\phi J}$ that commutes with the evolution generated by $\mathcal{L}$: $e^{\mathcal{L}t}(U^\dagger \rho U)=U^\dagger e^{\mathcal{L}t}(\rho) U$. 
The inverse of these implications is, however, not true. A conserved quantity is not necessarily a strong symmetry. Neither is a symmetry that globally commutes with the Lindbladian, as this does not imply it should commute with all of its terms separately. Such a symmetry $U$ is called weak. An example of weak symmetries is given by operators $U$ satisfying $\left[H,U\right]=0$ and $U F_l U^\dagger = e^{i\phi_l} F_l$, for any phases $\phi_l$. There is also no generic relationship between a weak symmetry and a conserved quantity.  

The practical consequences for Lindblad dynamics are as follows. First, since conserved quantities do not decay along the evolution, they must give rise to a degenerate steady-state subspace. Second, weak symmetries block-diagonalise $\mathcal{L}$, but these blocks do not necessarily contain a stationary eigenmode.
Weak symmetries have additional consequences on trajectories in that they commute with $H_{\text{eff}}$, so there exist quantities that are conserved between quantum jumps, but may be changed when a jump occurs.

\section{Model}

The system we employ to illustrate these general principles and their effect on SPT phases is the cluster model: a one-dimensional spin-1/2 chain of length $N$ with open boundary conditions.
In the following, we denote Pauli matrices on site $l$ of the chain as $\sigma^x_l=X_l$ and likewise for $Y_l, Z_l$. The unitary evolution is governed by the cluster Hamiltonian $H_0$ and a perturbation term $H_{xx}$, $H=H_0+H_{xx}$:
\begin{equation}
	H_0=J\sum_{l=2}^{N-1} K_l,\;\; H_{xx}=V_{xx}\sum_{l=1}^{N-1} X_{l}X_{l+1},
	\label{eq:Hzxz+HV}
\end{equation}
with $K_l=Z_{l-1}X_l Z_{l+1}$ - the cluster operators.
The model $H_0$ is invariant under spin-flip symmetries on odd and even sites, denoted $G_{o/e}$, that generate a global $\mathds{Z}_2\cross\mathds{Z}_2$ symmetry, which leads to the existence of trivial and non-trivial SPT phases~\cite{Bahri2015,PhysRevB.104.014424,PhysRevLett.125.200506,Son_2011}.
Since all terms in $H_0$ commute with each other, it is frustration-free and readily solvable, as shown in Appendix~\ref{appendix:cluster}.

In the following, we consider the impact of symmetries defining SPT phases becoming either strong, weak or neither when the system interacts with an external environment.
For this purpose, we choose various types of jump operators $F_i$ to model differing system-bath interactions.
The most general jumps can be written as a sum of Pauli strings. For example, we may consider the lowering operator along the $x$-basis, $F_l=S_{xl}^{-}=(Z_l+iY_l)/2$, which preserves the $G=\mathds{Z}_2\cross\mathds{Z}_2$ symmetry as a weak one, but destroys the edge-mode and the bulk-$ZXZ$ symmetries.

\begin{figure*}
	\centering
	\raisebox{-0.5\height}{\includegraphics{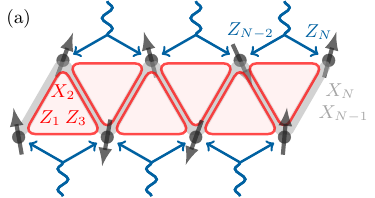}}\hspace{4mm}%
    \raisebox{-0.5\height}{\includegraphics{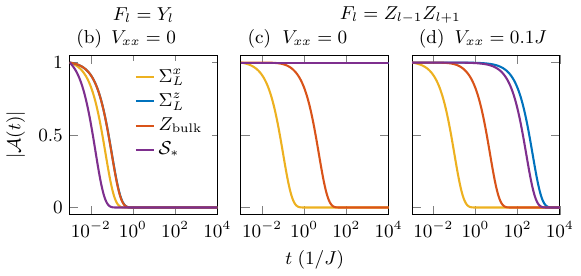}}%
    \caption{\label{fig:dynam}(a) Illustration of the model with three types of interactions. Red triangles: 3-spin cluster/$ZXZ$ interactions. Gray lines: 2-spin $XX$ perturbation. Blue wavy arrows: example of coupling of this system to an environment that acts as $ZIZ$, \textit{i.e.} jointly on two neighbouring spins on the odd or even sublattice. The odd and even sublattices are respectively the lower and upper legs of this ladder-like structure. (b)-(d) Auto-correlation of observables of interest against time: $\Sigma_L^{x,z}$ - components of the left edge mode of the cluster model, $Z_{\text{bulk}}$ - spin polarization of some site in the bulk, $\mathcal{S}_*$ - string order parameter defined in the text. Left: for a symmetry-breaking dissipation. Center and right: with the symmetry-preserving dissipation studied in this paper. All for $N=8, \kappa=2.5 J$, starting from an initial cluster state. In (b) and (c), the curve corresponding to $\Sigma_L^{z}$ is hidden below the curves of $Z_{\text{bulk}}$ and $\mathcal{S}_*$, respectively.}
\end{figure*}

Remarkably, we observe that the large class of models with jump operators that are each an individual Pauli string preserve all of the relevant symmetries of the model at least weakly. For example, the system with $F_l=Y_l$ has all of them as weak symmetries. 
Furthermore, it is also possible to promote some of them to strong symmetries, notably with $F_l=Z_{l-1}Z_{l+1} \; (l=2,3,...,N-1)$, $G$ and the edge modes components $Z_1, Z_N$ are strong. It is illustrated on \figref{fig:dynam}[a]. We develop a detailed understanding of the time-scales for the persistence of the local and global symmetries under the open system dynamics.

\subsection{Topological signatures in the closed system}

For later comparison to understand the effect of weakening or destroying these symmetries through bath interactions, we now summarize some of the topological properties they induce in the closed model. With open boundary conditions, the model has a ground state with a symmetric bulk, while the action of the symmetry on the edges is said to fractionalize. In the non-trivial SPT phase, it gives rise to spin-$\frac{1}{2}$ degrees of freedom localized on the edges of the chain and independent of the dynamics. For $H_0$, on the left boundary we have: $\Sigma^x_L=X_1 Z_2, \Sigma^y_L=Y_1 Z_2, \Sigma^z_L=Z_1$
and similarly on the right end. In this case, these are actually symmetries of the model (also called strong zero modes), so they exist not only in the ground state subspace but across the whole spectrum. 

When spin-flip-symmetry-preserving perturbations such as $H_{xx}$ are added, up to a critical strength, edge modes are guaranteed to exist in the ground state and their mutual anticommutation is protected (since the SPT phase is characterized by a discrete value, it cannot change smoothly). However, their explicit form changes, developing an exponential tail away from the edge. Beyond the critical value, the discrete invariant characterizing the projective representations of the $\mathds{Z}_2\cross \mathds{Z}_2$ changes abruptly, the phase becomes topologically trivial, and the edge modes are lost~\cite{PhysRevB.96.165124}. 

The possibility for these modes to exist, and hence the non-trivial SPT phase, can be detected by degeneracies in the entanglement spectrum~\cite{PhysRevB.81.064439,PhysRevB.102.235157}, which are four-fold in the cluster model.
The entanglement spectrum is the spectrum of the reduced density matrix of the state, obtained by partially tracing out a sub-region of the spin chain.
Intuitively, cutting the system in two pieces and discarding one half in order to compute their entanglement introduces ambiguity in the state due to the cutting of the global $G_{o/e}$ symmetries, creating a degenerate mixture of possible states on the remaining half of the system (see Appendix~\ref{app:cluster-degeneracy} for details). 
Later, this feature will provide a key probe of SPT order along quantum trajectories.

Different SPT phases can be distinguished by string order parameters that measure a global order~\cite{PhysRevB.86.125441}. 
For a symmetry acting on local sites with $u_i$, such that $U=\prod_{i} u_i$, and edge operators $O^{L/R}$, a general string order parameter is defined as 
\begin{equation}
     \mathcal{S}(U, O^L, O^R)= \lim_{k \to \infty} \left< O^L_1 \prod_{i=2}^{k-1} u_i O^R_k \right>.
\end{equation}
There exist selection rules following which suitable choices of $O^{L/R}$ allow to distinguish the different topological phases~\cite{PhysRevB.86.125441}. 
Here, we will observe that one example of such an operator exists in the stationary state of the studied system with a certain choice of jump operators.

%% file: 3_strong.tex
We now focus on the effects of strong symmetries that remain in the presence of environmental interactions, first demonstrating that for certain dissipators, a steady state manifold containing logical qubits may survive.
To understand the resilience of these qubits to perturbation in the thermodynamic limit, we will study the time it takes to relax to this steady state manifold in the unperturbed system, finding it remains finite at all system sizes.
Thus, for a sufficiently small perturbation, the steady state manifold will remain long-lived and maintain coherence (i.e. it becomes a metastable manifold), rather than becoming mixed with other degrees of freedom.
Further, while most perturbations would need to weaken with system size to maintain metastability, we will see that perturbations preserving the $\mathds{Z}_2\cross\mathds{Z}_2$ symmetry create a system-size independent metastable time-scale, thus preserving long-lived qubits in the thermodynamic limit even at non-zero strength.

\subsection{Steady-state manifold}

As discussed earlier, the most immediate consequence of strong symmetries is the presence of conserved quantities, and thus a degenerate steady state manifold.
We begin by describing the structure of this manifold and the topological features which survive in the unperturbed system $\mathcal{L}_0$ (\textit{i.e.} $\mathcal{L}$ with $V=0$), focusing primarily on $F_l=Z_{l-1}Z_{l+1}$ jump operators.
In this case, the dissipation is strongly symmetric with respect to the $\mathds{Z}_2\cross\mathds{Z}_2$ spin flips. 
Further, the edge mode operators $\Sigma^z_{L/R}$ are also preserved as a strong symmetry. 
In contrast, the $x$ and $y$-components of the edge modes are only weak symmetries and are therefore not guaranteed to affect the stationary state. 

In order to quantify how the information from the initial state survives along the time evolution we employ the autocorrelation of measurement outcomes of an observable $O$, as visible in the averaged Lindbladian dynamics. In general terms, it is the sum over all possible measurement outcomes - $\nu_i$ at initial time and $\nu_j$ after time $t$ - weighted by the probabilities of these outcomes to occur given an initial state $\ket{\psi_0}$ and some temporal dynamics: 
\begin{equation}
    \mathcal{A}(t)=\sum_{i,j} \nu_i \nu_j P(i|\psi_0) P(j|t,i,\psi_0).
\end{equation} 
A detailed derivation is presented in Appendix~\ref{appendix:autocorr}.

As expected from the remaining strong symmetries with $F_l=Z_{l-1}Z_{l+1}$ jumps, \figref{fig:dynam}[c] shows the edge operator $\Sigma^z_L$ and the string order $\mathcal{S}_* \equiv \mathcal{S}(G_oG_e,Y,Y)$ are conserved by the dynamics, while another component of the edge mode, $\Sigma^x_L$, and some $Z$ in the bulk decay. In contrast, as \figref{fig:dynam}[b] shows, $F_l=Y_l$ jumps make all these quantities decay rapidly and none of them is preserved in the steady state.

In total there are 16 conserved quantities with $F_l=Z_{l-1}Z_{l+1}$ jumps, generated by the odd and even spin-flip parities and by the left and right $z$-components of the edge modes. 
Remarkably, we note that the corresponding 16-dimensional steady-state subspace forms two Pauli groups, which means that it can be used to encode quantum information. The logical qubits are delocalized over odd or even sites of the lattice. In analogy to the edge modes of the closed system, we label them as odd and even logical qubits and they are as follows:
\begin{center}
	\begin{tabular}{ c | c }
		odd sites & even sites \\ \hline 
		$\mathds{1}$ & $\mathds{1}$ \\ 
		$G_o$ & $G_e$ \\ 
        $Z_1G_o$ & $G_eZ_N$ \\
		$Z_1$ & $Z_N$ \\ \hline
		$\shortparallel$ & $\shortparallel$ \\
		$\vec{\Sigma}_o$ & $\vec{\Sigma}_e$
	\end{tabular}
\end{center}
Therefore, despite the fact that the edge modes of the closed system dissipate and can not serve as quantum memory in trajectory-averaged behaviour, $F_l=Z_{l-1}Z_{l+1}$ jump operators preserve strong symmetries that allow us to identify two distinct qubits that form a decoherence-free subsystem~\cite{PhysRevA.63.042307,Lidar2003}. All other degrees of freedom decohere to the fully mixed state, so the steady state can be rewritten as
\begin{equation}
\rho_{ss}=\frac{1}{4}\left(\mathds{1}+ \sum_{i,j}d_{ij} \; \Sigma^i_o \otimes \Sigma^j_e \right)\otimes \frac{\mathds{1}}{2^{N-2}}, 
\end{equation}
where the coefficients $d_{ij}$ determine the joint state of the two qubits, encoding all surviving initial state information. The corresponding change of basis between the physical and logical qubits is detailed in Appendix~\ref{appendix:basis} and indicates the physical implementation of logical single- and two-qubit gates. 

We note that here $\mathcal{S}(G_oG_e,Y,Y)$ is conserved. 
Such string-order parameters can act as topological markers in mixed states, and imply computational power for MBQC~\cite{Raussendorf2022Oct}. 
In the context of MBQC, the string correlator measures the localizable entanglement, \textit{i.e.} the hidden correlations needed for a state to serve as a computational resource~\cite{PhysRevA.80.022316}.
It has been proven that string order survives at finite times~\cite{deGroot2022symmetryprotected} when dissipators leave the symmetry protecting the topological phase as a strong symmetry. 
The conservation of $\mathcal{S}(G_oG_e,Y,Y)$ here implies it is also possible for some to survive in the infinite time limit. 

In addition, the steady-state manifold can contain certain mixtures of cluster states.
Although the entanglement spectrum is in general not a good witness of quantum entanglement for mixed states, it can detect SPT order in cluster states, which have a 4-fold degenerate spectrum. 
As we show in Appendix~\ref{app:cluster-degeneracy}, mixtures of cluster states also possess a 4-fold degeneracy as a direct consequence of 4-fold degeneracies of individual cluster states.
Later we will show this 4-fold degeneracy is also maintained along the trajectory dynamics of individual pure states, when starting from a cluster state, due to the presence of weak symmetries.

\subsection{Dissipative gap}

\begin{figure}[h] 
	\centering
	\includegraphics{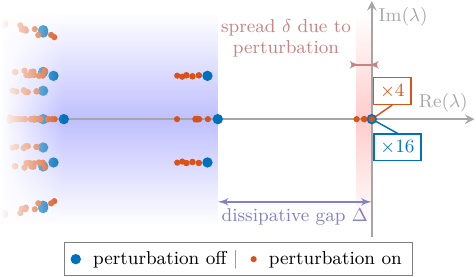}
    \caption{\label{fig:spec}Sketch of the low-lying spectrum of the model with $ZIZ$ jump operators. Without the perturbation $H_{xx}$, the spectrum is highly degenerate, in particular the steady-state has multiplicity $16$. The next levels are separated by the dissipative gap $\Delta$. With the perturbation on, some degeneracies are lifted and the steady state spreads by $\delta$, thus acquiring a finite lifetime, except for $4$ states remaining at $0$.}
\end{figure}%

While the presence of stationary qubits is a desirable quality, for practical purposes this would ideally be resilient to perturbations.
Further, it is necessary to understand the system size and parameter dependence of this resilience in order to tune the system for maximum performance.
To this end, we will begin by studying the low-lying spectral properties of the Lindbladian $\mathcal{L}_0$, describing long-time dynamics just prior to the steady-state. 
The spectral properties of Lindbladian superoperators have been investigated, using techniques such as integrability~\cite{Prosen2008Apr,PhysRevLett.122.010401,PhysRevE.102.062210,PhysRevLett.117.137202,SciPostPhys.8.3.044} and random matrix theory~\cite{PhysRevLett.123.140403,Can2019Nov,PhysRevLett.123.234103,Sa2020Jul}  -- in the following, we will take an approach similar to the former.
We will consider the influence of perturbations on the steady-state manifold, relative to the timescales of the unperturbed system, to understand the impact of dissipation on the coherence time of these qubits.

To study the low-lying spectrum, we first note the presence of low-dimensional subspaces under the repeated action of the Lindbladian, i.e. Krylov subspaces, the smallest of which enable a closed form solution of a small number of eigenvalues.
These smallest spaces originate from dynamically local operators, i.e. those locally differing from the steady-state manifold, such as a stationary operator multiplied by a small number of Pauli operators.
Earlier works have shown that local operators may lead to the long time-scale dynamics in Lindbladian evolution~\cite{PhysRevLett.124.100604,PhysRevResearch.3.023190}, suggesting that these spaces may correspond to the low-lying spectrum in our case.
In the following, we will identify which of these smallest spaces are minimally influenced by the dissipative jump operators that determine relaxation.
We argue that this minimal dissipation, combined with their locality, suggests these spaces determine the longest timescales in the system and thus contain the dissipative gap, which we verify numerically up to $N=16$.

In more detail, recall that the unperturbed Hamiltonian $H_0$ is defined as a sum of mutually commuting local operators $K_l=Z_{l-1}X_l Z_{l+1}$; likewise, the dissipation consists of mutually commuting local operators $F_l=Z_{l-1}Z_{l+1}$. 
We may thus consider the root state to be a Pauli string that either:
\begin{enumerate}[label=(\alph*)]
    \item does not undergo unitary evolution (i.e. commutes with $H_0$) but experiences non-zero dissipation due to a set of $m$ jump operators,
    \item annihilated by the action of of all jump-operators except $n$ ``active'' sites where the state anti-commutes with the associated $K_l$ terms from $H_0$,
\end{enumerate}
with $m=n=0$ corresponding to the identity steady state, and small $m$ or $n$ corresponding to operators we refer to as dynamically local.
All the operators in category (a) are eigenstates of $\mathcal{L}_0$ with eigenvalues $-4m\kappa$, since $\mathcal{D}(O)\propto O$ for any Pauli string $O$ due to $F_l$ themselves being Pauli strings. 
However, an operator $O$ in category (b) is not an eigenstate itself but is coupled by the unitary dynamics to $2^n -1$ other operators.
The latter possess the same anti-commutation properties with the $K_l$ operators as the initial operator $O$, but they are no longer zero under action by the jump operators. 
Increasing $n$ leads to bigger invariant subspaces containing operators affected by more jump operators. 
We thus focus on low $n$ cases, i.e. those originating from dynamically local operators, which we expect to produce subspaces with the weakest coupling to the environment, and thus the slowest dynamics.

We will now describe the cases of $n=1$ and $n=2$ anti-commuting sites in detail, finding the lowest non-zero eigenvalue contained in these subspaces across the parameter space.
Then, in the absence of a complete mathematical proof, we have verified that this eigenvalue and accompanying heuristic reasoning matches precisely with the exact numerical solution of the dissipative gap for system sizes up to $N=16$, suggesting it is valid in the thermodynamic limit (see Appendix~\ref{appendix:excited}).

\subsubsection{One active site}
\begin{figure}[h] 
	\centering
	\includegraphics{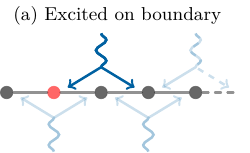}
	\includegraphics{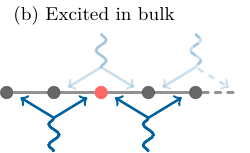}
    \caption{\label{fig:krylov}Excited active site (in red) affected by a different number of $ZIZ$ jump operators depending on its position. Relevant dissipators are drawn in dark blue, irrelevant ones -- in light blue.}
\end{figure}

In scenario (b) with $n=1$, we start with a Pauli string $A$ such that $\mathcal{D}(A)=0$, but also such that $\left[K_l,A\right]=0$ for all $l$ except at some site $p$, which we call an active site, where the Pauli string anti-commutes with the local cluster operator, $K_pA=-AK_p$. 
This implies that effectively only a single term of the unitary evolution contributes to the master equation, so that
\begin{equation}
    \mathcal{L}_0(A) = \mathcal{U}_0(A) = -2iJ K_p A \equiv 2J B.
\end{equation}
In that way, we have defined state $B\equiv -iK_p A$, in which site $p$ has been excited by the action of $K_p$. As it has the same commutation relationship with $K_l$'s as $A$, the action of $\mathcal{L}_0$ on $B$, due to the unitary part on site $p$ yields
\begin{equation}
    \mathcal{U}_0(B) = -2iJ K_p B = -2J K_p^2 A = -2J A,
\end{equation}
because $K_p^2=1$.
On the other hand, the jumps $F_{p\pm 1}$ have a dissipating effect on $B$ as they do not commute with the $K_p$ present in the definition of $B$. 
The dissipator thus yields 
\begin{equation}
    \mathcal{D}(B)=-4\alpha \kappa B,
\end{equation}
where: $\alpha=1$ if the active site $p$ resides next to one of the boundaries ($p=2$ or $N-1$), where only one of the jump operators can dissipate it; and $\alpha=2$ if the active site is further into the bulk ($p=3,4,...,N-2$), where two jump operators can dissipate it.
This is illustrated in \figref{fig:krylov}. 

We have in this manner shown that $\{A,B\}$ is a subspace invariant under the action of $\mathcal{L}_0=\mathcal{U}_0+\mathcal{D}$. 
Its lower eigenvalue as a function of $\kappa/J$ is 
\begin{equation}
\lambda_1(\alpha)=-2\alpha\kappa+2\sqrt{\alpha^2\kappa^2-J^2}. \label{eq:delta1}
\end{equation}
For $\kappa/J < 1/\sqrt{3}$, the $\alpha=1$ case is closest to the steady state, and beyond this point, there is a level crossing with the state corresponding to $\alpha=2$ case. 
We remark that these kind of invariant subspaces have an exceptional point (EP) at $\kappa/J=1$ for $\alpha=1$ or $\kappa/J=0.5$ for $\alpha=2$, at which both eigenvalues and eigenvectors coalesce with one another.
These exceptional degeneracies are intensely studied - in the context of Lindblad and non-Hermitian physics with parity-time symmetry breaking - for their mathematical properites and practical physical consequences~\cite{El-Ganainy2018Jan, Ozdemir2019Aug, Ashida2020Jul, RevModPhys.93.015005}.

The lowest-lying eigenstates corresponding to these levels are expressed as
\begin{equation}
W_p(\alpha) = \left[\mathds{1} +i\frac{\lambda_1(\alpha)}{2J} K_p\right]A. \label{eq:Wp}
\end{equation}
It is a superposition of two operators from the invariant subspace, but since all of them can be expressed in terms of a single root state $A$ and the cluster operators $K_l$, it takes this remarkably simple form. 
Concrete examples are:
\begin{align}
    \alpha=1\: :&\:\: A=Z_{2} \:\: \textrm{or}\:\: A=Z_{N-1},\nonumber\\
    \alpha=2\: :&\:\: A=Z_{i}, \:\: i=3,...,N-2.
\end{align}
In these cases it is immediately clear that the corresponding anti-commuting $K_p$ is located at the site $p$ where the single $Z$ is located. 
In one strong symmetry sector there would be $2$ such local operators near the boundaries and $N-4$ in the bulk.
Note that the operators described here are not physical density matrices by themselves, but a density matrix can contain such dissipating components. 
They are a useful way of organizing the eigenmodes of the Lindbladian and interpreting the relaxation dynamics.

\subsubsection{Two active sites}
Continuing in this vein but starting from a non-dissipated Pauli string $A$ that now anti-commutes with the cluster operators $K_l$ at two different sites $p$ and $q$ (corresponding to category (b) with $n=2$). 
We show that such operators can also give rise to excitations that determine the dissipative gap in some $\kappa /J$ regimes, becoming longer lived than the $n=1$ case.

A similar reasoning leads to building a four-dimensional invariant subspace $\{A,B,C,D\}$ given by
\begin{align}
    B&\equiv -i K_p A,\nonumber\\ C&\equiv -i K_q A,\\ D&\equiv - K_p K_q A. \nonumber
\end{align}
Now, the action of the dissipator depends on the positions of the active sites $p$ and $q$ relative to both the boundaries and also to each other, and whether these sites are excited, i.e. whether the operator contains the corresponding factor of $K_p$ and/or $K_q$. 

\begin{figure}[h] 
	\centering
	\includegraphics{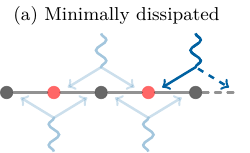}
	\includegraphics{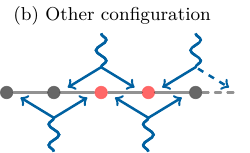}
    \caption{\label{fig:krylov2}Excited active sites (in red) affected by a different number of $ZIZ$ jump operators depending on their position on the chain and with respect to each other. Relevant dissipators are drawn in dark blue, irrelevant ones in light blue. Case (a) illustrates that if next-to-nearest neighbours are both excited (i.e. state $D=- K_p K_q A$ with $p=q\pm2$), the dissipator acting on both of them is neutralized.}
\end{figure}

By inspecting all possibilities for active sites $p$ and $q$, the closest states to the steady state are those for which the active sites $p$ and $q$ are next-to-nearest neighbours, with one located next to one of the boundaries. 
These conditions minimize the effect of dissipation for two reasons.
First, the active site next to the boundary can only trigger the influence of at most one jump operator, as in the single excited site case illustrated in \figref{fig:krylov}[a], due to the lack of jump operators centred on the boundary sites.
Secondly, when the excited active sites are next-to-nearest neighbours, $p=q\pm2$, the corresponding operator $D=- K_p K_q A$ is not influenced by the jump $F_{(p+q)/2}$, further reducing the effect of the bath, as shown on \figref{fig:krylov2}[a].
In contrast, changing the distance between the active sites allows this operator to be affected by additional dissipative channels, as exemplified in \figref{fig:krylov2}[b].

This kind of operator exists only for $N\geq 6$ because of the spacing needed for these excitations, and each strong symmetry sector contains only two such operators, one at each of the boundaries: $p=2,q=4$ or $p=N-3,q=N-1$.
Concretely, they are for example generated from $A=Z_2Z_4$ or $Z_{N-3}Z_{N-1}$.

The eigenstates are, similar to Eq.~\eqref{eq:Wp}, of the form 
\begin{equation}
    W_{p,q} = \left[\mathds{1} + u K_p + v K_q + w K_p K_q\right]A,\label{eq:Wpq}
\end{equation}
with coefficients $u,v,w$ depending on $\kappa$ and $J$, as detailed in Appendix~\ref{appendix:excited}. 
The lowest-lying eigenvalue is, in this case,
\begin{equation}
\lambda_2=-4\kappa + 2\sqrt{2}\sqrt{\sqrt{J^4 - J^2 \kappa^2 + \kappa^4} - J^2 + \kappa^2}.\label{eq:delta2}
\end{equation}
Interestingly, contrary to the singly dissipated states for Eq.~\eqref{eq:delta1}, the doubly dissipated case corresponding to Eq.~\eqref{eq:delta2} do not possess any EPs.
It seems that EPs are avoided due to the interplay between the active sites which is absent when the active sites are further apart, for example on the two boundaries, and do not see each other via a dissipator, in which case EPs do exist.

\subsubsection{Summary and numerical verification}
\begin{figure}[h] 
	\centering
	\includegraphics{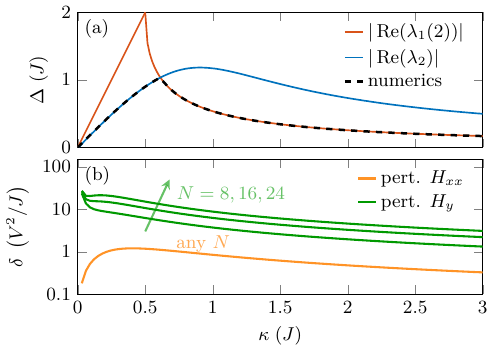}
    \caption{\label{fig:gapvskappa}(a) Dependence on $\kappa /J$ of the dissipative gap, in absolute value, of the unperturbed model $\mathcal{L}_0$. The gap has been verified numerically for system sizes from $N=6$ up to $N=16$, and this exact functional form is conjectured to hold for any $N$. (b) Dependence on $\kappa /J$ of the displacement $\delta$ of the steady-state spectrum, to second order of PT, as a consequence of the $H_{xx}$ and the symmetry-breaking $H_y$ perturbations. In the units, $V$ is $V_{xx}$ or $V_y$ correspondingly.}
\end{figure}

For small values of $\kappa/J$, below $\kappa/J=\sqrt{3/8}\approx 0.612$, it turns out the two active site eigenvalue $\lambda_2$ of Eq.~\eqref{eq:delta2} is lower than the single active site eigenvalue $\lambda_1(\alpha)$ of Eq.~\eqref{eq:delta1}, as shown in \figref{fig:gapvskappa}[a].
Since $\lambda_1(1)$ was only lower than $\lambda_1(2)$ for $\kappa/J < 1/\sqrt{3}$ it does not correspond to the lowest eigenvalue for any parameters.
Our reasoning thus postulates that the dissipative gap is given by $\Delta=\min(|\Re{\lambda_1(2)}|,|\Re{\lambda_2}|)$. It has a cusp due to levels crossing at $\kappa/J=\sqrt{3/8}$. A similar form of the gap was observed in Ref.~\cite{PhysRevB.99.174303}. 

That the dissipative gap corresponds to $\Delta$ is verified by exact numerical solutions up to $N=16$ in Appendix~\ref{appendix:excited}.
We have verified numerically that the two types of eigenstates studied above are sufficient to correctly capture the spectral gap. 
In fact, exact numerical diagonalization for $N$ up to 16, which is also drawn in \figref{fig:gapvskappa}[a], is in perfect agreement with the analytical prediction. 
Moreover, the reasoning we outlined above is independent of system size $N$ (for $N\geq 6$) as it relies on local excitations, which the numerical observations in Appendix~\ref{appendix:excited} also suggest.
In this way, the steady-state subspace is shown to be separated from the rest of the spectrum by a dissipative gap that persists in the thermodynamic limit.

\subsection{Resilience to perturbations}
We now turn to the question of stability to perturbation of the decoherence-free subsystem. 
We are interested in quantifying the newly acquired lifetime of the degenerate steady-state subspace, given by $\delta$ depicted on \figref{fig:spec}.
Under conditions that this timescale remains long relative to the dissipative gap $\Delta$ of the unperturbed system, the steady-state manifold will become a metastable manifold, and coherence will still be preserved up to long times.

Metastability is known to occur generically when breaking strong symmetries with a sufficiently weak perturbation \cite{Macieszczak2016,Rose2016,Macieszczak2021b}.
However, the size of a perturbative parameter which qualifies as weak depends both on the original dissipative gap and the system size dependence of the perturbation.
If the perturbative gap becomes comparable to the dissipative gap, the dynamics will entangle the qubit degrees of freedom with the remainder of the system, destroying the preservation of information on short timescales.

We first consider the case of a perturbation which preserves the global $\mathds{Z}_2\cross\mathds{Z}_2$ symmetry, for example, the $H_{xx}$ defined in Eq.~\eqref{eq:Hzxz+HV}.
As such, the symmetry operators $G_{o/e}$ are still conserved; in contrast, $Z_1$ and $Z_N$ are not conserved, since they no longer commute with the Hamiltonian.
However, as seen in \figref{fig:dynam}[d], they decay with a longer timescale than other degrees of freedom for small values of the perturbing field. 

We investigate the lifetime of the broken symmetries using second-order degenerate perturbation theory (PT) for open quantum systems. The Lindbladian superoperator can now be written as 
\begin{equation}
    \mathcal{L}=\mathcal{L}_0+\mathcal{V},\text{ with } \mathcal{V}=-i\left[H_{xx},\;\cdot\; \right].
\end{equation}
After splitting the system into slow and fast subspaces $\mathcal{P}$ and $\mathcal{Q}$ - the former containing the steady-state space of $\mathcal{L}_0$ and the latter all the rest - PT up to second order yields~\cite{PhysRevA.86.012126}:
\begin{align}
	L_1^{\text{eff}}&=P\mathcal{V}P, \\
	L_2^{\text{eff}}&= - P\mathcal{V}Q\mathcal{L}_0^{-1}Q\mathcal{V}P. \label{eq:L2eff}
\end{align}
$P$ and $Q$ are projectors onto the slow and fast spaces, and $\mathcal{L}_0^{-1}\equiv (Q\mathcal{L}_0 Q)^{-1}$.

The first order vanishes because the perturbation does not affect the steady-state space directly. 
At second order, steady states are coupled to either one or two subspaces of $\mathcal{Q}$ that are connected by the action of $\mathcal{L}_0$.
We note that the perturbation does not connect any stationary states to the same subspaces, and as such does not mix steady states among themselves: hence $L_2^{\text{eff}}$ is diagonal.
Further, we note that this diagonal is real, as it can be shown that when considering a Hermitian basis (in this case, the steady state manifold) and a Hamiltonian perturbation, $L_2^\text{eff}$ must be real.

The subspaces the perturbation connects each stationary state to happen to be small, a property originating in an algebraic structure present in the model, to be elaborated upon in future work.
For example, $\Sigma_o^z = Z_1$ from the $(-,+)$ sector is coupled only with the following four-dimensional connected subspace of $\mathcal{Q}$: $\{Y_1X_2, X_1Z_3, Y_1Y_2X_3Z_4, X_1Z_2Y_3Z_4\}$.
This structure enables exact solutions, demonstrated in Appendix~\ref{appendix:PT} for $L_{2,Z_1}^{\text{eff}}$, the diagonal element of $L_2^{\text{eff}}$ corresponding to steady-state $Z_1$.

The elements of all 16 basis elements of the steady-state subspace turn out to be closely related.
To see this, we may separate them into four sectors of the $G_{o/e}$ symmetries, labeled by $(\pm , \pm)$. 
The $(+,+)$ sector is unaffected by the perturbation, because they are eigenmatrices of the remaining strong symmetries.
Within either the $(+,-)$ or $(-,+)$ sectors, diagonal entries of $L_2^{\text{eff}}$ are the same since the associated steady states are related by a multiple of the $G_{o/e}$ symmetry operator.
Further, the entries for the $(+,-)$ and $(-,+)$ sectors are also the same as each other due to reflection symmetry.
Meanwhile, entries are twice as large for the $(-,-)$ sector: the perturbation couples these to exactly twice as many subspaces of $\mathcal{L}_0$, which are related by reflection symmetry and thus induce the same contribution to the diagonal entry. 
The $(-,-)$ sector thus determines the maximal spread $\delta=\max(\text{diag}\left(-L_2^{\text{eff}}\right))$.

The $\kappa /J$ dependence of $\delta$ is shown in \figref{fig:gapvskappa}[b]. 
Its inverse provides the timescale of the decay acquired due to the perturbation.
We extract the expected lifetime of, for example, $\Sigma^z_L=Z_1$ at $\kappa/J=2.5$ and find it to be $ t^* = 1/ |L_2^{\text{eff}}| \approx 5.149 (V_{xx}/J)^{-2} J^{-1}$.
This agrees perfectly with the exact dynamics of \figref{fig:dynam}[d] which displays a lifetime of $514.2 J^{-1}$ for $V_{xx}=0.1J$, suggesting that second order perturbation theory is sufficient to capture the correct dynamics even for a relatively large perturbing field.

We note importantly that when the Hamiltonian perturbation preserves the $\mathds{Z}_2\cross\mathds{Z}_2$ symmetry, $\delta$ is independent of system size $N$ in this case.
This is due to such a perturbation connecting each steady state to a non-extensive number of subspaces, each with dimensions which do not scale with system size.
Since the dissipative gap of $\mathcal{L}_0$ also appears to be independent of system size, we see that it is possible for a finite (i.e. non-zero) perturbation to retain a metastable manifold in the thermodynamic limit.
As such, in the presence of a symmetry-preserving perturbation, the system possesses a symmetry-protected metastable manifold consisting of the logical qubits: while these decohere in the infinite time limit, they remain coherent across large timescales.

In contrast, a perturbation that breaks the spin-flip $\mathds{Z}_2\times \mathds{Z}_2$ symmetry will affect the bulk $X$ strings of the steady-state components.
This will cause a spread $\delta$ that scales extensively with system size, since such steady states will be coupled to extensively many subspaces of $\mathcal{L}_0$. 
For example, consider perturbing the Hamiltonian with
\begin{equation}
    H_{y}=V_y \sum_{l=1}^N Y_l.
\end{equation}
The exact expression for $\delta$ as a function of $N$ is given in Appendix~\ref{appendix:PT} and is drawn on \figref{fig:gapvskappa}[b] for $N=8,16,24$.
We observe that the highest rate at which a steady state decays predicted by second-order PT now increases linearly with system size.
Thus in the thermodynamic limit, the model with $H_y$ shows non-perturbative behaviour that cannot be captured by PT: however, the behaviour of PT at finite sizes suggests the logical qubits will interact with other degrees of freedom increasingly as system size grows for any finite (i.e. non-zero) perturbation, causing them to lose coherence.

\subsection{Summary of strongly-symmetric qubits}
To conclude this section, our analysis has shown that the cluster model with the $ZIZ$ dissipators displays a stationary state manifold with a range of interesting features.
In particular, this includes the presence of two non-local logical qubits, separated from other degrees of freedom by a finite gap in the thermodynamic limit. 
We have then observed that with global symmetry-preserving perturbations, these stationary features can remain long-lived compared to the rest of the system, residing in a symmetry-protected metastable manifold.
In contrast, symmetry-breaking perturbations likely cause increasingly rapid decoherence as system size grows.
An important part of the perturbation theory analysis is played by the structure in the spectrum, which will be the subject of a future article.

%% file: 4_weak.tex
Strong symmetries have significant effects on the steady-state manifold of Lindbladian dynamics. 
Weak symmetries, on the other hand, do not, but they do have non-trivial consequences for quantum trajectories.
In general, there exists a particular representation of jump operators under which all weak symmetry operators commute with $H_0$ and gain a phase under commutation with the jumps \cite{Macieszczak2021a}.
With Pauli jump operators $F_l$, our model is naturally in this representation: all symmetry operators $U$ of the cluster model $H_0$ remain at least weak symmetries, and as Pauli strings they thus satisfy $U F_l U^\dagger=\pm F_l$.
We now explore the impact these properties have on trajectory dynamics within the model.

\subsection{Weakly conserved observables}
The most immediate consequences of weak symmetries comes from their commutation with $F_l^{\dagger} F_l$ for all $l$, and therefore with $H_{\text{eff}}$: weak symmetry operators are thus conserved quantities between the quantum jumps.
We demonstrate this using three types of jump operators:
\begin{itemize}
    \item $F_l=Z_{l-1}Z_{l+1}$ for $l=2,...,N-2$: as discussed earlier, this leaves the $G_{o/e}$ spin flip symmetries and edge operators $Z_{1}$ and $Z_{N}$ as strong symmetries but makes the remaining symmetries -- $X_1Z_2$, $Z_{N-1}X_N$ and $K_l=Z_{l-1}X_lZ_{l+1}$ for $l=2,...,N-2$ -- as weak symmetries.
    \item $F_l=Y_l$ for $l=1,...,N$: this makes all symmetries of $H_0$ weak.
    \item $F_l=S_{xl}^-$ for $l=1,...,N$: a non-Pauli string jump operator, this breaks all symmetries aside from $G_{o/e}$, which are left weak.
\end{itemize}

\begin{figure}[h] 
	\centering
    \includegraphics{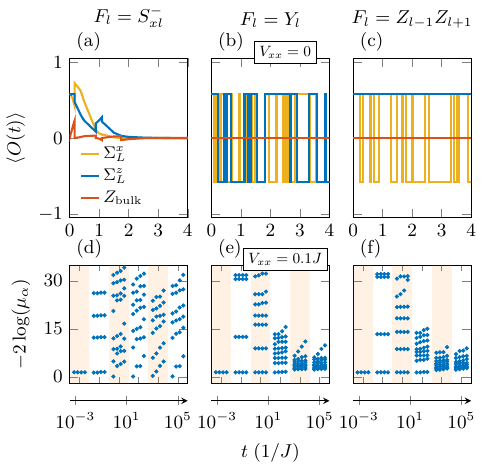}
    \caption{\label{fig:traj}(a)-(c) Evolution of observables of interest along a single quantum trajectory for different jump operators, without perturbation. The initial state is a cluster state in the ground state of $H_0$, with the left edge mode having equal expectation of all three components and the right one along $+z$ direction. (d)-(i) Entanglement spectrum of one trajectory at a few instants of the dynamics, with perturbation. $\mu_\alpha$ are the Schmidt coefficients of the state. Parameters: $N=10, \kappa=2.5 J$. The initial state is the cluster state in the ground state of $H_0$, in the $(+,+)$ sector of the $\mathds{Z}_2\cross\mathds{Z}_2$ symmetry.}
\end{figure}

\figref{fig:traj}[a-c] compares the behaviour of one trajectory with the different jumps - it shows the dynamics of the expectation value of three observables: two components of the left edge mode, $\Sigma^x_L=X_1 Z_2$ and $\Sigma^z_L=Z_1$, and a $Z$ operator in the bulk. 
\figref{fig:dynam}[c] indicated that on average $F_l=Z_{l-1}Z_{l+1}$ caused $\Sigma^x_L$ to decay, while \figref{fig:dynam}[b] that $F_l=Y_l$ caused both $\Sigma^x_L$ and $\Sigma^Z_L$ to decay. 
However, looking at individual realizations of the time evolution with the same jumps in \figref{fig:traj}[b-c], we observe that memory of the initial state is not entirely lost, as expected from weak symmetries.
Instead, jumps cause components of the edge mode to be reflected stochastically, but they are conserved between these jumps and only attain a small subset of values. 
In contrast, the dynamics induced by $S_{xl}^-$ destroys any information in the initial state even within individual trajectories, as shown in \figref{fig:traj}[a].

\subsection{Entanglement spectrum degeneracy}
In addition to weak edge symmetries, the $F_l=Y_l$ and $F_l=Z_{l-1}Z_{l+1}$ jump operators leave all the interior $K_l=Z_{l-1}X_lZ_{l+1}$ as weak symmetries.
States that are simultaneous eigenstates of all these operators, along with being eigenstates of either $\{G_o, G_e\}$ or $\{\Sigma^x_L, \Sigma^x_R\}$, are called cluster states.
Due to the weak symmetry properties, the action of any Pauli string jump operator on a cluster state thus remains a cluster state, albeit potentially a different one due to a sign change of the state's eigenvalues under some symmetry operators.
Both choices of using $\{G_o, G_e\}$ and $\{\Sigma^x_L, \Sigma^x_R\}$ are detailed in Appendix~\ref{appendix:cluster}.

A key signature of SPT order in these states is degeneracy of the entanglement spectrum: this is 4-fold degenerate for $\{G_o, G_e\}$, and 2-fold degenerate for $\{\Sigma^x_L, \Sigma^x_R\}$ (see Appendix~\ref{app:cluster-degeneracy} for details).
Since an initial cluster state remains a cluster state along a quantum trajectory in this model, the entanglement spectrum keeps its original four-fold degeneracy along the trajectory in the case of Pauli-string jumps, while it does not otherwise.

When we reintroduce the $H_{xx}$ Hamiltonian perturbation, \figref{fig:traj}[e-f] shows that the degeneracy of the entanglement spectrum is also lifted for the Pauli-string jumps, although \figref{fig:traj}[d] shows it remains much longer lived compared to trajectories with non-Pauli $S_{xl}^-$ jumps.
We note that these properties also induce similar degeneracy properties in the mixed-state evolution due to the Lindblad equation: a mixture of cluster states possesses the same degeneracy properties as the individual cluster states (see Appendix~\ref{app:cluster-degeneracy} for details), and since the Lindblad evolution describes the average of trajectory evolutions, an initial mixture of cluster states remains a mixture of cluster states at all times.

\begin{figure}[ht] 
	\centering
    \includegraphics{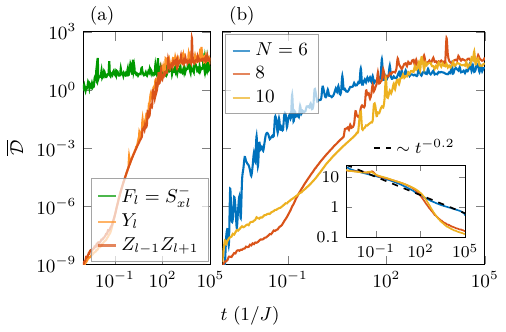}
    \caption{\label{fig:degen}Extent of the four biggest Schmidt values of the state in a trajectory, $\mathcal{D}=\frac{\log\mu_1 - \log\mu_4}{\log\mu_4 - \log\mu_5}$, characterizing the four-fold degeneracy of the spectrum, normalized by the gap to the next levels, averaged over trajectory realisations with randomly chosen initial cluster state. (a) For three different jump operators, at $N=8, \kappa=2.5 J, V_{xx}=0.1 J$. (b) Dependence on system size for $F_l=Z_{l-1}Z_{l+1}$ and $\kappa=2.5 J, V_{xx}=0.1 J$. Inset: evolution of the gap $(\log\mu_4 - \log\mu_5)$ for different system sizes.}
\end{figure}%

To quantify the evolution of degeneracy in the entanglement spectrum, we define
\begin{equation}
\mathcal{D}=\frac{\log\mu_1 - \log\mu_4}{\log\mu_4 - \log\mu_5},    
\end{equation}
where $\mu_\alpha$ are Schmidt coefficients of the state, ordered from largest to smallest. 
It measures the ratio between the extent of the first four levels of the spectrum and the gap separating them from the fifth one. 
\figref{fig:degen}[a] displays this quantity averaged over trajectory realizations, which demonstrates the prolonged life of the degeneracy in the case of Pauli string jumps.
Remarkably, this lifetime increases with the system size, as shown in \figref{fig:degen}[b].
The inset therein shows the evolution of the gap, $\log\mu_4 - \log\mu_5$, for the three system sizes. 
This gap protects the degeneracy of the dominant Schmidt coefficients, and hence the topological character of the state. 
For a chain of 6 spins, it roughly follows a power law decay $\sim t^{-0.2}$, which also holds for 8 and 10 spins, but the behaviour goes through a change at $t\approx 100 J^{-1}$, which is when $\overline{\mathcal{D}}$ starts to plateau.

\section{Weakly symmetric qubits}
We now focus specifically on the edge qubit operators, e.g. for the left edge $\Sigma^x_L=X_1 Z_2$, $\Sigma^z_L=Z_1$, and the corresponding $\Sigma^y_L=Y_1 Z_2$.
As discussed previously, these are weak symmetries whose expectations are conserved between jumps and change by at most a sign under the action of a Pauli string jump operator.
Noting that the state of the edge qubit is uniquely defined by these expectations, a vector $\left[\langle X_1Z_2\rangle_0, \langle Y_1Z_2\rangle_0, \langle Z_1\rangle_0\right]$ on the Bloch sphere, with $F_l=Y_l$ jump operators we find:
\begin{center}
    \includegraphics{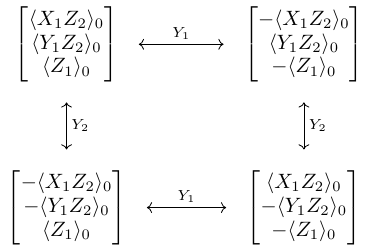}
\end{center}
with all other jump operators leaving the left edge qubit unchanged.

Each Pauli jump therefore acts as a $\pi$-rotation about the $x,y$ or $z$ axis in the Bloch sphere, as illustrated in \figref{fig:bloch}[a].
If it commutes with component $x$ but anti-commutes with $z$, it flips the state onto $\ket{\psi_0^x}$, and if the inverse is true - onto $\ket{\psi_0^z}$. 
If it anti-commutes with both $x$ and $z$ components, the state becomes $\ket{\psi_0^y}$. 

This introduces a dichotomy between two types of qubits: those residing in the steady state due to strong symmetries of the Lindblad master equation, which we call \textit{strong qubits}; and those supported by the weak symmetries, lost in the Lindbladian dynamics but visible in individual quantum trajectories, which we call \textit{weak qubits}.
We note in particular that the weak qubit operators are local, and thus may be easier to manipulate than the global strong qubit operators considered earlier.
For this to be useful, however, we need to demonstrate how to recover the initial states of the weak qubits.

\subsection{Recovering quantum information from weakly symmetric qubits}
It is possible to use the weak symmetry properties to return the weakly symmetric qubit to its original state.
One approach is to count the occurrences of jumps which affect the weakly symmetric qubits, e.g. the number of occurrences of $Y_1$ and $Y_2$ for the left qubit with $Y$ jumps. 
The initial information can then be retrieved by applying an appropriate operation at the end of the trajectory: flipping the spin back from $\ket{\psi^{x,y,z}_0}$ to $\ket{\psi_0}$ if needed. 
We note that this operation can be any operator performing the required rotation of $\pi$ around $x,y$ or $z$ axes on the weak qubits. 
Notably, it can be local even if the jump operators or the weak qubit itself is delocalized over many sites. 

A second approach is more closely related to standard error correction techniques, with the weak symmetry operators $K_l$ viewed as stabilizers \cite{Gottesman1997}.
As noted above, provided a state is an eigenstate of these operators at the start, Pauli string jump operators will leave the state as a potentially distinct eigenstate.
If there is a one-to-one correspondence between the effect of jumps on the weakly symmetric qubits and the effect of jumps on the eigenvalues of the $K_l$ operators, then measurements of these operators can uniquely determine what error has occurred on the weakly symmetric qubits and thus informs what correction needs to be applied.

The applicability of the error correction based approach thus depends on the noise channels considered, since each distinct combination of jumps must be distinguishable through measurements of the $K_l$ stabilizer operators.
In practice, the experimental feasibility of measuring these stabilizers, or alternatively observing which jumps occur, would be a key factor in choosing which protocol to implement.
However, we note that observing jumps would work for any set of noise channels, and is more clearly resistant to Hamiltonian perturbation.
We thus focus on this protocol in the next section.

\subsection{Resilience to perturbation}
\begin{figure}[h] 
	\centering
    \includegraphics{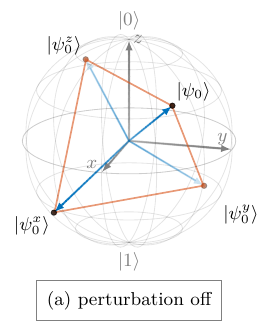}%
    \includegraphics{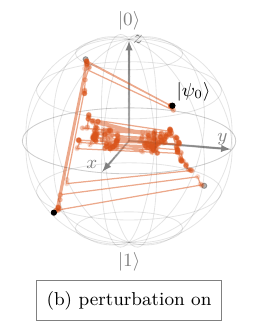}%
    \caption{\label{fig:bloch}Schematic Bloch sphere of the weak qubit during a quantum trajectory evolution. (a) Without perturbation it alternately occupies 4 states as there is no evolution between jumps, and the latter can only produce $\pi$-rotations about the Bloch sphere axes. Restoration procedure is exact - it consists in monitoring the number of flips that appear in each direction and applying a restoring flip at measurement time. (b) Perturbation makes it slowly deviate from those 4 states, hence loosing in fidelity until reaching and oscillating around the origin. Restoration can still undo the effect of jumps to some extent, such that the weak qubit is made long-lived. Parameters for illustration: $N=8$, $\kappa=0.1J$, $V_{xx}=0.1J$, trajectory shown up to $t=316J^{-1}$.}
\end{figure}

In the presence of the $H_{xx}$ Hamiltonian perturbation ($V=0.1J$) the edge qubits evolve between jumps, due to the breakage of weak symmetries, as illustrated in \figref{fig:bloch}[b]. 
This causes the state to progressively drift away from the initial state on a time scale dictated by the perturbation strength, eventually reaching the maximally mixed state due to entanglement with bulk degrees of freedom. 
Before this relaxation, the trajectory stays in the vicinity of $\ket{\psi_0}$ and $\ket{\psi^{x,y,z}_0}$, since the $Y_1$ and $Y_2$ jumps still act as rotations. 
By applying the error-mitigation protocol the effects of the jumps can be mitigated on average, hence enhancing the lifetime of the qubit, where the lifetime is dictated by the perturbation and not the dissipation. 

\begin{figure}[h] 
	\centering
    \includegraphics{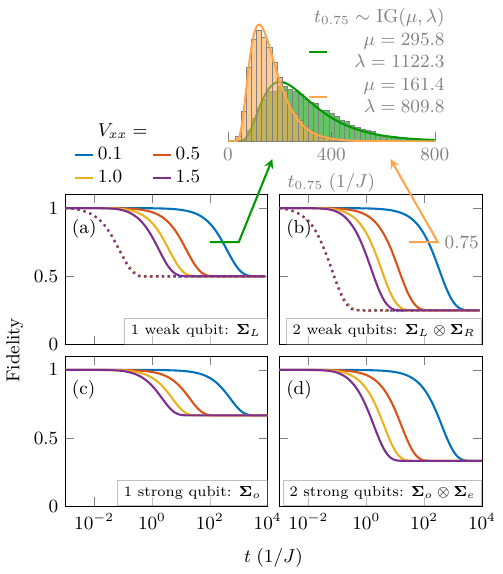}
	\caption{\label{fig:fidelity}Evolution of the fidelity of (a) the left edge mode, and (b) the two edge qubits together, for $Y_l$ jumps at $N=8, \kappa=2.5 J$ and different values of the perturbation strength $V_{xx}$. Solid lines show the fidelity when the restoring procedure is applied, dotted lines show it without doing anything and they all overlap. The inset histograms show the distribution of times at which the fidelity for single trajectories goes below $0.75$ for the first time. Averaged over $50000$ trajectory realisations. (c) and (d) show the fidelity for the strong qubits, i.e. the ones existing in the steady-state of the Lindbladian dynamics with $Z_{l-1}Z_{l+1}$ jumps. For all curves, the initial state is a cluster state with all three components of the left edge mode nonzero, in the ground state of $H_0$, although no dependence on the initial state has been observed. Also, no apparent dependence on system size.}
\end{figure}%

We quantify the memory of the encoded qubit in terms of fidelity, measured by the distance between mixed states $\rho$ and $\sigma$ as ~\cite{Jozsa1994}
\begin{equation}
F(\rho,\sigma)=\left(\Tr{\sqrt{\sqrt{\rho} \sigma \sqrt{\rho}}}\right)^2.
\end{equation}
This is presented for states belonging to one and two weak edge qubits in \figref{fig:fidelity}[a-b]; for comparison, the fidelity of the steady-state strong qubits, discussed in Sec~\ref{sec:ss}, is displayed in \figref{fig:fidelity}[c-d].
The initial state is the ground state of $H_0$ with the left edge mode $\Vec{\Sigma}_L$ in a configuration with all three components having equal expectation value, and the right one, $\Vec{\Sigma}_R$, pointing in the $+z$ direction.

When the error-mitigation protocol is not applied, fidelity decays equally rapidly for all values of perturbative field strengths $V$, indicated by the dashed lines in \figref{fig:fidelity}[a-b]. 
In the absence of perturbation, the protocol proposed above preserves the information indefinitely, hence the fidelity remains $1$ at all times. 
In the presence of perturbation $H_{xx}$, the protocol extends the lifetime of the weak qubits as shown in \figref{fig:fidelity}[a-b], becoming comparable to the lifetimes of the strong qubits as shown in \figref{fig:fidelity}[c-d].
All lifetimes reduce with increasing $V$ as shown in \figref{fig:fidelity}[a-d].
We also note that the above results are independent of system size or initial state, likely due to the same algebraic properties that led to system size independent time scales under perturbation for the strong qubits.

In the ensemble of trajectories, the timescale at which the fidelity of the error-mitigated state drops below $0.75$ is a fluctuating measure of the lifetime of the logical qubit. The distribution of lifetimes, as shown in the histograms of \figref{fig:fidelity}, is well approximated by an inverse Gaussian distribution, $t_{0.75}\sim \text{ IG}(\mu,\lambda)$.
Its probability density function (pdf) is
\begin{equation}
    f(t|\mu,\lambda)=\sqrt{\frac{\lambda}{2\pi t^3}} \exp[\frac{-\lambda(t -\mu)^2}{2\mu^2 t}], t>0
\end{equation}
where $\mu>0$ and $\lambda>0$ are the mean and shape parameter.

We remark that the inverse Gaussian characterizes the first passage time distribution for Brownian motion with a drift~\cite{ROY1968191}, which thus offers a possible explanation for the behaviour of the first passage time $t_{0.75}$. 
The evolution of the fidelity could therefore be imagined as resembling a Wiener process due to the quantum dissipative jumps and subject to drift due to the perturbation. 
This is however only an intuitive picture: a full understanding would require considering the dynamics of the fidelity induced by the underlying jump process of the quantum state~\cite{jumpprocess} and taking into account the contribution of the perturbation to the fidelity. 

\subsection{General weakly symmetric qubits}
Through the above illustration we have demonstrated a general principle: weak symmetries in quantum trajectories allow for information contained in the initial state to be recovered, provided environmental interactions can be sufficiently monitored.
When these weak symmetries form an $\mathfrak{su}(2)$ algebra, or a product of multiple $\mathfrak{su}(2)$ algebras, this information consists of qubits and can thus encode quantum information.

In the context of quantum information in open systems, it is therefore worth exploring models beyond those hosting qubits in the steady state of a Lindbladian. 
By looking at the weak symmetries and their implications for trajectories, a broader class of models may be able to preserve quantum information.

%% file: 9_conclusion.tex
In this work, we have uncovered a rich set of dynamical phenomena associated with strong and weak discrete symmetries in open quantum systems. 
We characterize the entanglement and topological properties of a one-dimensional spin-$1/2$ chain in the presence of various forms of dissipation. The unperturbed Hamiltonian involves the $3$-spin $ZXZ$ interaction, often called the cluster model, which has a global $Z_2 \times Z_2$ symmetry. This symmetry being strong results in a degenerate steady state manifold. While the degeneracy is lifted on perturbing the Hamiltonian, the manifold can remain metastable in the thermodynamic limit, as long as the symmetry is not broken. 
Our results point towards a general mechanism for \textit{symmetry-protected metastability}, potentially enabling the design of models with coherent behaviours that survive to times inversely proportional to the square of the perturbation strength, which can be much longer than the relaxation time of the unperturbed dynamics.

The robust metastability in this setup originates from the low-lying spectrum of the Lindbladian which we have studied using the algebraic structure of the Hamiltonian and jump operators. 
Specifically, the Hamiltonian we considered is a sum of elements of a stabilizer group and the jump operators commute or anti-commute with these stabilizers.
This structure allows for the exact solution of the unperturbed low-lying spectrum, while also dampening the effect of symmetry-preserving perturbations on the spectrum.
The structure is intricately connected to the symmetries of the model, which have further implications for the entire spectrum and constrain the dynamics of observables. 
A framework for studying this class of algebraic structures in interacting Lindbladians could shed light on their integrable properties, a topic of considerable interest.

The stationary manifold of the Lindbladian contains states that exhibit a four-fold degenerate entanglement spectrum, a signature of the entanglement and symmetry-protected topological (SPT) order associated with closed systems, remarkably surviving in the presence of dissipation.
For dissipation acting as a Pauli string on the state, the degeneracy of the entanglement spectrum is maintained throughout the evolution when the system is initialized in a cluster state.
We have demonstrated that these signatures originate from the effect of weak symmetries on quantum trajectories, an unravelling of the Lindbladian dynamics into the noisy dynamics of a pure state.
We see that the SPT order can be preserved along the individual realizations of the trajectories, leading to the survival of the entanglement spectrum degeneracy.
Our observations suggest a key perspective for understanding topological physics in open quantum systems through the lens of weak symmetries and trajectories. 
Further, while we have focused on a particular class of jumps (Pauli strings), more general forms of dissipation could also respect weak symmetries and hence lead to similarly interesting behaviours.
In the future, it would be interesting to further classify the impact of dissipation on SPT phases using quantum trajectories, as has been initiated in Ref.~\cite{PhysRevResearch.4.023036} for another class of models.

Finally, we have shown that the symmetry-protected metastable manifold originating from the strong symmetries harbours logical qubits, albeit with a non-locality that potentially makes them hard to manipulate. 
We have also demonstrated that weak symmetries can enable local edge qubits which retain quantum information within individual quantum trajectories, invisible to the trajectory-averaged Lindbladian evolution.
The state of an edge qubit can be restored through a protocol based on monitoring environmental interactions of a small section of the spin chain, enhancing the qubit's lifetime, which we further demonstrate is resilient to perturbation.
Our results motivate trajectories as a new avenue for controlling entanglement in dissipative systems and stabilizing novel states in noisy intermediate-scale quantum devices.

Realizing the fundamental properties of this family of models could be potentially achieved in a variety of physical settings.
Recent experiments have realized SPT phases in isolated interacting models using cold atoms in an optical lattice~\cite{doi:10.1126/science.aav9105,Sompet2022}. Engineered dissipation is also an experimental tool increasingly prevalent for the preparation of desired states and the study of non-equilibrium quantum phenomena~
\cite{Mi2023Apr}. 
Three-qubit unitary interactions, an essential ingredient of our theory, have been engineered recently~\cite{Kim2022Jul,Warren2023May} in quantum circuits of superconducting qubits which also allow for controlled dissipative processes and measurements to achieve a desired steady-state~\cite{PhysRevA.88.023849,PhysRevA.88.032317,Kapit_2017,PhysRevLett.116.240503}. Importantly, these devices can be operated in the regime of monitored dynamics to study the statistical properties of quantum trajectories~\cite{PhysRevA.77.012112, murch2013, hatridge2013,  Roch2014}.
Third, NISQ (Noisy Intermediate Scale Quantum) devices can also host cluster states, as has been demonstrated in Ref.~\cite{PhysRevLett.125.120502}.
This could be used to probe results presented here and further develop our understanding of topological phases in open quantum systems.

\begin{acknowledgements}
We would like to thank Christopher Turner, Juan Garrahan, and Rosario Fazio for their valuable insights and discussions. 
M.H.S.\ gratefully acknowledges  financial support from EPSRC
(Grants No. EP/S019669/1 and EP/V026496/1).
A.P.\ and D.C.R. are funded by the European Research Council (ERC) under the EU’s Horizon 2020 research and innovation program (Grant Agreement No.~853368).
D.P.\ is funded by the UCL Graduate Research Scholarship. 
The authors acknowledge the use of the UCL Myriad High Performance Computing Facility (Myriad@UCL), and associated support services, in the completion of this work.
\end{acknowledgements}

%% file: A1_mathdetails.tex
\section{Autocorrelation of measurement outcomes}\label{appendix:autocorr}

The autocorrelation of measurement outcomes of an observable $O$ is used in the main text (see \figref{fig:dynam}) to probe survivial of initial information throughout the evolution of the open quantum system. 

The observable can be decomposed as $O=\sum_i \nu_i P_i$, with $\nu_i$'s being its eigenvalues, \textit{i.e.} the possible measurment outcomes, and $P_i$'s - the projectors on the corresponding eigenspaces. 

$\mathcal{A}(t)$ quantifies the correlation between the results of two measurements $\nu_i$, $\nu_j$ at an initial and final time, mediated by a given temporal change in the state labeled by $r$, averaged over samples:
\begin{equation}
    \mathcal{A}(t) = \sum_{i,j,r} \nu_i \nu_j P(i|\psi_0) P(r|i,t,\psi_0) P(j|t,r,i,\psi_0) ,
\end{equation}
where $P(i|\psi_0) = \Tr{P_i\ket{\psi_0}\bra{\psi_0}}$ is the probability of outcome $i$ happening given the initial state $\ket{\psi_0}$, $P(r|i,t,\psi_0)$ is the probability that the temporal evolution $E_r(t)$ occurs given the initial measurement outcome $i$ on $\ket{\psi_0}$, 
\begin{align}
    P(r|i,t,\psi_0) = \frac{\Tr{E_r(t) P_i \ket{\psi_0}\bra{\psi_0} P_i E_r^\dagger (t)}}{\Tr{P_i\ket{\psi_0}\bra{\psi_0}}},
\end{align}
and $P(j|t,r,i,\psi_0)$ is the probability of measuring outcome $j$ given initial measurement $i$ and evolution $r$
\begin{equation}
    P(j|t,r,i,\psi_0) = \frac{\Tr{P_j E_r(t) P_i \ket{\psi_0}\bra{\psi_0} P_i E_r^\dagger (t)}}{\Tr{E_r(t)P_i\ket{\psi_0}\bra{\psi_0}P_i E_r^\dagger (t)}}. 
\end{equation}

The sum over $j$ can be performed to recover $O$, and putting all the probability definitions together leads to
\begin{equation}
    \mathcal{A}(t)=\sum_{i,r} \nu_i \Tr{O E_r(t) P_i \ket{\psi_0}\bra{\psi_0} P_i E_r^\dagger (t)}.
\end{equation}

Further, carrying out the sum over all possible trajectories $r$, the time evolution reduces to the Lindbladian one:
\begin{equation}
    \mathcal{A}(t)
    = \sum_i \nu_i P(i|\psi_0) \Tr{O e^{\mathcal{L}t}(\rho_i(0))} \label{eq:AL},
\end{equation}
where $\rho_i(0)$ is the properly normalized density matrix after the first measurement at time $0$.
Finally, using the linearity of the Lindbladian evolution, we may evolve an unphysical matrix $\tilde{\rho}(0)=\sum_i\nu_i P(i|\psi_0) \rho_i(0)$
\begin{equation}
    \mathcal{A}(t)
    = \Tr{O e^{\mathcal{L}t}(\tilde{\rho}(0))} \label{eq:AL2}.
\end{equation}

\section{Cluster states}\label{appendix:cluster}

The cluster Hamiltonian $H_0$ from~\eqref{eq:Hzxz+HV} is frustration-free, that is all its terms mutually commute. As it is shown here, its eigenstates are simply build out of the ones of each of the cluster terms. Additionally, due to the open boundary conditions, some degrees of freedom at the edges remain free, so that each energy level is degenerate with four-fold multiplicity, accommodating for the four possible states of the edges.

First, we can rewrite the problem in the basis of clusters on each site by observing that the following operators form spin algebras: 
\begin{equation}
    \begin{aligned}
        \Tilde{Z}_l &= K_l = Z_{l-1}X_{l}Z_{l+1}\\
        \Tilde{X}_l &= Z_l\\
        \Tilde{Y}_l &= -Z_{l-1}Y_{l}Z_{l+1}
    \end{aligned}
\end{equation}
This is valid in the bulk, \textit{i.e.} for $l=2,...,N-1$. On the edges, they become $\Tilde{Z}_1=X_{1}Z_{2}, \Tilde{X}_1=Z_{1}, \Tilde{Y}_1=-Y_{1}Z_{2}$ and $\Tilde{Z}_N=Z_{N-1}X_{N}, \Tilde{X}_N=Z_{N}, \Tilde{Y}_N=-Z_{N-1}Y_{N}$. Analogously to spins, $\Tilde{Z}_l$ gives the cluster eigenvalue at $l$, while $\Tilde{X}_l$ and $\Tilde{Y}_l$ flip it up to some phase. We can easily see that these operators respect spin commutation and anti-commutation relations and define a new on-site, local basis for a chain of clusters. Note that in the main text, in order to match the notation from the literature, the edge modes are written as $\Tilde{Z}_{1/N}\equiv \Sigma^x_{L/R}$ and $\Tilde{X}_{1/N}\equiv \Sigma^z_{L/R}$.

This maps the model onto a new spin chain with the Hamiltonian $H_0=\sum_{l=2}^{N-1} \Tilde{Z}_l$.
The two clusters at the boundaries are strong zero modes, as they do not appear in the Hamiltonian. 
Hence the eigenstates can be written simply in terms of the eigenstates of the $2^{N-2}$ $\Tilde{Z}_l$'s in the bulk: $\ket{+}$ and $\ket{-}$, with eigenvalues $+1$ and $-1$ respectively. Each of these will be four-fold degenerate, as the edge clusters are free. This thus forms a complete and orthonormal basis of the Hilbert space, labeled by the eigenvalue on each site, which is denoted by the set $\{\pm_l\}$. In the original basis, these states can be obtained by applying corresponding cluster raising and lowering operators $\Tilde{S}_l^{\pm} = (\Tilde{X}_l \pm i\Tilde{Y}_l)/2$ on the state of all spins up $ \ket{\uparrow}^{\otimes N} $:
\begin{equation}\label{eq:noGcluster}
\ket{ \{\pm_l\} } = \sqrt{2^N} \left( \prod_{l=1}^N \Tilde{S}_l^{\pm_l} \right) \ket{\uparrow}^{\otimes N}.
\end{equation}

With this formulation of the problem, the action of the jumps on the cluster basis states is readily understood. 
In this new basis, for example the $Y_l$ jumps act like $-\Tilde{X}_{l-1}\Tilde{Y}_{l}\Tilde{X}_{l+1}$ in the bulk ($l=2,...,N-1$) and $-\Tilde{Y}_{1}\Tilde{X}_{2}$ and $-\Tilde{X}_{N-1}\Tilde{Y}_{N}$ on the edges. Their action is then to flip the cluster eigenstates on two or three adjacent sites. But when starting from a cluster state, along a single trajectory the state always remains a cluster state. In particular, only $Y_1$ and $Y_2$ flip the left edge mode. The consequences of that are illustrated on \figref{fig:traj}[a-c] and explored further in Section~\ref{sec:traj} of the main text. This is the case for any jump operator that is a Pauli string, as it will also be a Pauli string in the new basis, \textit{i.e.} it will preserve the cluster structure of the state. On the contrary, jumps that are sums of Pauli strings create generate superposition of states. In the main text we illustrate it with $S_{xl}^-=(Z_l+iY_l)/2$ and see that it indeed destroys cluster states and their short-range entanglement along a single trajectory. 

Alternatively, on \figref{fig:traj}[d-f] we consider states that have four-fold degenerate entanglement spectra. These are cluster states as well, that is eigenstates of $H_0$, but also eigenstates of the spin-flip symmetries. They arise if, in the preceding reasoning, we take the spin-flip symmetries instead of the edge modes in order to construct them. The previous basis allows to explore different configurations of the edge modes, while these states allow to more clearly characterize the topology through 4-fold degeneracy in the entanglement spectrum.

They can be defined as~\cite{PhysRevB.104.014424}
\begin{equation}
\ket{ \{\pm_l\}, g_o, g_e } = \sqrt{2^N} \left( \prod_{l=2}^{N-1} \Tilde{S}_l^{\pm_l} \right) P_o^{g_o'} P_e^{g_e'}  \ket{\uparrow}^{\otimes N}, \label{eq:Gcluster}
\end{equation}
where $P^\pm_{o/e}=(1\pm G_{o/e})/2$ are projectors onto the sectors of the spin-flip symmetries with eigenvalues $g_{o/e}=\pm 1$, and $g'_{o/e}=g_{o/e}^{N/2-1}=\pm 1$. The action of the Pauli-string jump operators on them is also still preserving the cluster nature of the state, but can also change the spin-flip-symmetry sector. 
Bulk operators $\Tilde{X}$ and $\Tilde{Y}$ act like previously by flipping the cluster eigenvalue, but they also flip $g_o$ or $g_e$ if they act on an odd or even site. Bulk $\Tilde{Z}$ returns the cluster eigenvalue. The action on the edges is more complicated. First, $\Tilde{X}_{1/N}=Z_{1/N}$ commute through everything in~\eqref{eq:Gcluster} except for $G_{o/e}$ respectively, of which they change the eigenvalue. Then, noting that $\Tilde{X_1}G_e = (\Tilde{Z}_2\Tilde{Z}_4...\Tilde{Z}_{N-2})\Tilde{Z}_N$ we find 
\begin{equation}
\Tilde{Z}_N\ket{\{\pm_l\}, g_o, g_e} = (-1)^{p_e}g_e\ket{\{\pm_l\}, -g_o, g_e}
\end{equation}
with $p_e=\sum_{l=1}^{N/2-1} \Tilde{Z}_l$. Analogous reasoning holds for the other edge. 

In conclusion, that demonstrates why Pauli-string jump operators also preserve the cluster states in this representation in terms of the $\mathds{Z}_2\cross \mathds{Z}_2$ symmetry operators, as used in \figref{fig:traj}[d-f].

\section{Entanglement spectrum of cluster states and their mixtures}\label{app:cluster-degeneracy}
Each cluster state (either Eq.~\eqref{eq:noGcluster} or Eq.~\eqref{eq:Gcluster}) is an eigenstates of a set of $N$ commuting parity operators $O_l$. We may write their density matrix as a normalized product of the projection operators onto the corresponding eigenstates:
\begin{equation}
    \rho_{\kappa} = \prod_{l=1}^N\frac{I+(-1)^{\kappa_l} O_l}{2},
\end{equation}
where $\kappa=[\kappa_i]_{i=1}^N$ with $\kappa\in\{0,1\}$ enumerate the cluster states.
Such a state is said to be stabilized by a set of $N$ stabilizers $K_l$, such that $K_l\ket{\psi}=\ket{\psi}$ and $[K_l,K_{l'}]=0$, where each $K_l$ is a Pauli string. 
In this case $K^{\kappa_l}_l=(-1)^{\kappa_l} O_l$.
We may thus write
\begin{equation}
    \rho_\kappa = \prod\frac{I+K^{\kappa_l}_l}{2} = \frac{1}{2^N}\sum_{\sigma_1^N} \prod_l (K^{\kappa_l}_l)^{\sigma_l},
\end{equation}
where $\sigma_l\in{0,1}$ and the sum over $\sigma_1^N=[\sigma_i]_{i=1}^N$ runs over all values.

The partial trace of this density matrix thus consists of a sum over partial traces of the Pauli strings $\prod_l (K^{\kappa_l}_l)^{\sigma_l}$.
These traces will fall into two classes of values for each Pauli string: they will be $0$ if the string has non-identity support on the traced part of the system; or they will be equal to themselves scaled by one over the Hilbert space dimension of the scaled space, e.g. $1/2^M$ for tracing out $M$ sites. We will specifically consider a contiguous set of $M$ sites starting from site $1$.

First, let us consider the case with stabilizers $K^{\kappa_l}_l=(-1)^{\kappa_l} Z_{l-1}X_lZ_{l+1}$ for $l=2,...,N-1$, $K^{\kappa_1}_1=(-1)^{\kappa_1} X_1Z_2$, and $K^{\kappa_N}_N=(-1)^{\kappa_N} Z_{N-1}X_N$, leading (up to a phase factor) to
\begin{equation}
    \prod_l (K^{\kappa_l}_l)^{\sigma_l} \propto X_1^{\sigma_1}Z_1^{\sigma_{2}}\left(\prod_{l=2}^{N-1}Z_l^{\sigma_{l-1}}X_l^{\sigma_l}Z_l^{\sigma_{l+1}}\right)Z_N^{\sigma_{N-1}}X_N^{\sigma_N}.
\end{equation}
Requiring site $1$ to be the identity leads to $\sigma_1=0$, $\sigma_2=0$.
Requiring site $l$ to be identity requires $\sigma_l=0$ and $\sigma_{l-1}=\sigma_{l+1}$.
Thus, requiring all sites $l=1,...,M$ to be identity requires $\sigma_l=0$ for $l=1,...,M+1$.
This leaves $2^{N-M - 1}$ remaining non-zero contributions to the partially traced density matrix
\begin{equation}
    \Tr_{1:M}\left(\rho_\kappa\right) = \frac{1}{2^{N-M}}\sum_{\sigma_{M+2}^N} \prod_{l=M+2}^N (K^{\kappa_l}_l)^{\sigma_l}.
\end{equation}

Next, consider the case with stabilizers $K^{\kappa_l}_l=(-1)^{\kappa_l} Z_{l-1}X_lZ_{l+1}$ for $l=2,...,N-1$, $K^{\kappa_1}_1=(-1)^{\kappa_1} G_o$, and $K^{\kappa_N}_N=(-1)^{\kappa_N} G_e$.
Using notation $\sigma_0 = \sigma_N$ we have
\begin{align}
    \prod_l (K^{\kappa_l}_l)^{\sigma_l} \propto& X_1^{\sigma_1}Z_1^{\sigma_{2}} X_2^{\sigma_{2}}Z_2^{\sigma_3}X_2^{\sigma_N} \nonumber\\
    &\times \left(\prod_{l=3}^{N-2}Z_l^{\sigma_{l-1}}X_l^{\sigma_l}Z_l^{\sigma_{l+1}}X_L^{\sigma_{l~(\textrm{mod}~2)}}\right) \nonumber\\
    &\times Z_{N-1}^{\sigma_{N-2}}X_{N-1}^{\sigma_{N-1}}X_{N-1}^{\sigma_1} Z_N^{\sigma_{N-1}}X_N^{\sigma_N}.
\end{align}
Starting again at site $1$, we find that for identity both $\sigma_1$ and $\sigma_2$ must be $0$.
Next, for site $2$ we find $\sigma_3$ and $\sigma_N$ must therefore also be $0$.
Requiring $2<l\leq M$ to be identity we thus find that since both $\sigma_1$ and $\sigma_N$ are $0$, we must have $\sigma_l=0$ and $\sigma_{l-1}=\sigma_{l+1}$.
We therefore have $\sigma_l=0$ for $l=1,...,M$, in addition to $\sigma_N=0$, leaving us with $2^{N-M-2}$ contributions to the partially traced density matrix, a factor of $2$ less than previously, giving
\begin{equation}\label{eq:cluster-partial-trace-flip}
    \Tr_{1:M}\left(\rho_\kappa\right) = \frac{1}{2^{N-M}}\sum_{\sigma_{M+2}{N-1}} \prod_{l=M+2}^{N-1} (K^{\kappa_l}_l)^{\sigma_l}.
\end{equation}

In order to demonstrate that these partially traced density matrices contain degenerate eigenvalues, we calculate their square
\begin{equation}
    \Tr_{1:M}\left(\rho_\kappa\right)^2 = \frac{1}{2^{2(N-M)}}\sum_{\sigma,\sigma'} \prod_{l} (K^{\kappa_l}_l)^{\sigma_l}(K^{\kappa_l}_l)^{\sigma'_l}.
\end{equation}
Noting that any set of stabilizers forms an Abelian group $\mathcal{G}$, by group axioms the left action of any element of that group forms an invertible function, and is thus a one-to-one mapping of the group.
We may therefore rewrite the square as
\begin{equation}
    \Tr_{1:M}\left(\rho_\kappa\right)^2 = \frac{1}{2^{2(N-M)}}\sum_{\sigma,\sigma'} \prod_{l} (K^{\kappa_l}_l)^{\sigma'_l},
\end{equation}
where we have reordered the sum after performing the action of each $(K^{\kappa_l}_l)^{\sigma_l}$, and thus
\begin{align}
    \Tr_{1:M}\left(\rho_\kappa\right)^2 =& \frac{|\mathcal{G}|}{2^{2(N-M)}}\sum_{\sigma} \prod_{l} (K^{\kappa_l}_l)^{\sigma_l}, \nonumber\\
    =& \frac{|\mathcal{G}|}{2^{N-M}}\Tr_{1:M}\left(\rho_\kappa\right)
\end{align}
where $|\mathcal{G}|$ is the number of elements in the group, i.e. $2$ to the number of stabilizers that generate it.

Since the densities power is equal to itself up to a scaling factor, this implies all of its eigenvalues are identical.
Further, since the trace of the density must be $1$, for $D$ eigenvalues of value $\lambda$ we have $D\lambda=1$ and from the purity we have
\begin{equation}
    D\lambda^2=\frac{|\mathcal{G}|D\lambda}{2^{N-M}},
\end{equation}
thus $\lambda=\frac{|\mathcal{G}|}{2^{N-M}}$ and $D=\frac{2^{N-M}}{|\mathcal{G}|}$.

For the two cases of stabilizers discussed previously: when the edge modes are used as stabilizers, $|\mathcal{G}|=2^{N-M-1}$, $\lambda=1/2$ and $D=2$; when the spin flip symmetries are used, $|\mathcal{G}|=2^{N-M-2}$, $\lambda=1/4$ and $D=4$.

Now consider a mixture of cluster density matrices,
\begin{equation}\label{eq:cluster-mixture}
    \rho = \sum_\kappa p_\kappa \rho_\kappa,
\end{equation}
for which we will study the entanglement spectrum via
\begin{equation}\label{eq:cluster-mixture-es}
    \Tr_{1:M}\left(\rho\right) = \sum_\kappa p_\kappa \Tr_{1:M}\left(\rho_\kappa\right).
\end{equation}
Noting that we may perform the sum in the partially traced cluster densities of e.g. Eq.~\eqref{eq:cluster-partial-trace-flip} to return to
\begin{equation}
    \Tr_{1:M}\left(\rho_\kappa\right) = \frac{1}{2^{N-M}}\prod_lI+K^{\kappa_l}_l,
\end{equation}
where the product is over the remaining stabilizers, e.g. $l$ runs over $M+2,...,N-1$ for the case with flip symmetry stabilizers and $M+2,...,N$ for the case with edge mode stabilizers.

Further, noting that 
\begin{equation}
    (I+K^a_l)(I+K^b_l)/4=\delta_{ab}(I+K^a)/2,
\end{equation}
we have
\begin{equation}
    \Tr_{1:M}\left(\rho_\kappa\right)\Tr_{1:M}\left(\rho_\tau\right)=\frac{|\mathcal{G}|\prod_l\delta_{\kappa_l\tau_l}}{2^{N-M}}\Tr_{1:M}\left(\rho_\kappa\right),
\end{equation}
implying due to hermiticity that these partial traces have disjoint supports if their signatures are distinct on the remaining part of the label.
In other words, each has non-zero eigenvalues with corresponding eigenvectors that are orthogonal to all eigenvectors with non-zero eigenvalues from another with a distinct signature, and thus contribute distinct eigenvalues.

In conclusion, we see that a classical mixture of cluster states such as Eq.~\eqref{eq:cluster-mixture} must have an entanglement spectrum with eigenvalues that are at least as degenerate as the degeneracy present in each individual cluster state, with further potential degeneracy originating from the probabilities in the mixture.

\section{Change of basis between physical and logical qubits}\label{appendix:basis}

The change of basis between the DFS form (decoherence-free subsystem) and the physical spin chain, highlighted in Sec.~\ref{sec:ss}, can conveniently be represented as a circuit of control $X$ gates. We will use the following rules to "push operators through" the CX gates: 

\begin{center}
\includegraphics{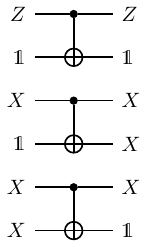}
\end{center}

It can be seen that a series of those gates transforms the chain of $X$ operators on the even or odd sites into a local $X$, which is the desired $\Sigma^x_{o/e}$ acting on the qubit degree of freedom.

The change of basis is thus $U=U_o U_e$, with 
\begin{equation}
	U_o=\prod_{i=1}^{N/2} \text{CX}_{1,2i-1}, \;\;\; U_e=\text{SWAP}_{N,2}\prod_{i=1}^{N/2} \text{CX}_{N,2i}, 
\end{equation}
where $\text{CX}_{i,j}$ is a CNOT/CX gate applied to the $i$ and $j$ spins and identity on every other spin, and $\text{SWAP}_{N,2}$ just swaps the vector spaces such that the two logical qubits are encoded on the first two vector spaces in the tensor product of the new basis. 

\section{Low-lying states}\label{appendix:excited}

In Sec.~\ref{sec:ss}, we give the form and eigenvalues of the "low-lying" states, those that establish the dissipative gap.

\begin{figure}[h] 
	\centering
	\includegraphics{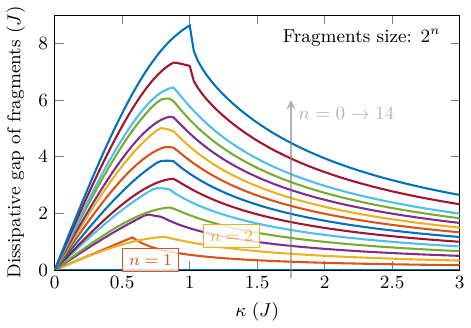}
    \caption{\label{fig:fraggapvskappa}The state space of the clean system ($V_{xx}=0$) of size $N$ divides into fragments of different dimensions. We regroup them into subsectors labeled by the size of the fragments, $2^n$ for $n=0\rightarrow N-2$. Here we show the dissipative gap of each such subsector for $N=16$. The dissipative gap of the whole system is governed by subsectors with $n=1$ and $2$.}
\end{figure}

As will be elucidated in further work, the system of $N$ spins has a fragmented state space with fragment sizes $2^n$, $n$ ranging from $0$ to $N-2$ and counting the number of sites at which the given states anti-commute with the cluster operators $K_l$. 
It can be seen readily by considering the Pauli basis, where each state naturally generates such a fragment when acted on with $\mathcal{L}_0$, as explained for $n=1$ and $2$ in Sec.~\ref{sec:ss}. 
The invariant subspaces uncovered there easily generalize to any $n$, hence partitioning the whole operator-space into an extensive number of fragments.
\figref{fig:fraggapvskappa} shows the dissipative gap among all the fragments of a particular dimension for $N=16$. 
More precisely, it is the gap separating the steady-state from the levels having the lowest real part, in absolute value, and belonging to that subsector of $2^n$-dimensional fragments. 
We observe that for the range of parameters displayed, fragments of size $2$ and $4$ are the ones setting the gap. 
It corresponds to the intuitive picture presented in the main text, according to which the gap is determined by those states that differ the least, by some "excitations", from the steady states, and hence are the least dissipated ones. 
Now, it can also be observed that increasing the system size from $N$ to $N+2$ does not modify the structure of the existing fragments, but demultiplies them and adds bigger, previously absent ones of size $2^N$. 
It follows by recursion that the dissipative gap always originates from $2$ and $4$-dimensional fragments, so we can be safe to limit our reasoning to them in the search of expressions for the gap, as has been done in Sec.~\ref{sec:ss}.

First, the $\lambda_1(\alpha)$ level of Eq.~\ref{eq:delta1} of the 2-dimensional fragments $\{A,B\}$ and the corresponding eigenmodes are straightforwardly obtained from the matrix representing the action of $\mathcal{L}_0$ therein
\begin{equation}
    \mathcal{L}^{(1)}_0(\alpha)=\begin{pmatrix} 0 & -2J \\ 2J & -4\alpha\kappa 
\end{pmatrix}.
\end{equation}

Second, to obtain the expression for $\lambda_2$, as in Eq.~\ref{eq:delta2}, we explore the different possibilities for the action of $\mathcal{L}_0$ in a 4-dimensional fragment $\{A,B,C,D\}$. We are interested in eigenvalues with the lowest possible real part (in absolute value), and we expect that to arise from a fragment subjected to the least possible amount of dissipation, hence a matrix with the smallest possible values on the diagonal. There are two possibilities, but the one among them with the lowest levels in all the parameter regimes is the following
\begin{equation}
\mathcal{L}^{(2)}_0=\begin{pmatrix} 0 & -2J & -2J & 0 \\ 2J & -4\kappa & 0 & -2J \\ 2J & 0 & -8\kappa & -2J \\ 0 & 2J & 2J & -4\kappa
\end{pmatrix}.
\end{equation}
Its lowest eigenvalue is~\eqref{eq:delta2} and the corresponding eigenmode~\eqref{eq:Wpq} is of the form $W = A + i u B + i v C - w D$ with 
\begin{equation}
    u=i\frac{(8\kappa + \lambda_2)\lambda_2}{4J(6\kappa + \lambda_2)},\; v=i\frac{(4\kappa + \lambda_2)\lambda_2}{4J(6\kappa + \lambda_2)},\; w=\frac{\lambda_2}{4\kappa + \lambda_2}.
\end{equation}
It corresponds to "excitations" such as intuitively described in the main text. 

\section{Perturbation theory}\label{appendix:PT}

In the first section below follow the details of the study of the $H_{xx}$ perturbation presented in Sec.~\ref{sec:ss}. This is an example of a perturbation introducing interactions but preserving the $\mathds{Z}_2\cross \mathds{Z}_2$ symmetry. We further contrast it to the case of a symmetry-breaking perturbation in the second subsection.

\subsection{$\mathds{Z}_2\cross \mathds{Z}_2$-symmetry-preserving perturbation}

In principle the perturbation could couple some steady-states among themselves. But this will not be the case here, which can be checked explicitly for the $X_lX_{l+1}$ terms of $H_{xx}$. Also, since they preserve the spin-flip symmetries, \textit{i.e.} they commute with $G_o$ and $G_e$, the $(+,+)$ sector is not affected. It will act similarly on the $(+,-)$ and $(-,+)$ sectors, and twice as much on the $(-,-)$ one, because it is a product of the latter two. Inside each one of those there are four states, but the perturbation has the same action on all of them. We thus choose to work on $\mathcal{P}=\{Z_1\}$, which through $\mathcal{V}$ is coupled only to the following invariant subspace of $\mathcal{Q}$: $\mathcal{Q'}=\{Y_1X_2, X_1Z_3, Y_1Y_2X_3Z_4, X_1Z_2Y_3Z_4\}$.
We refer to that diagonal entry of $L_2^{\text{eff}}$ as $L_{2,Z_1}^{\text{eff}}$.
In this fragment:

\begin{equation}
\begin{aligned}
\mathcal{L}_0^{-1}&=\begin{pmatrix}
-8\kappa & 2J & -2J & 0 \\
-2J & -4\kappa & 0 & 2J \\
2J & 0 & -8\kappa & 2J \\
0 & -2J & -2J & -4\kappa 
\end{pmatrix}^{-1} = S D^{-1} S^{-1},\\
D&=\text{diag}(\lambda_{-+}, \lambda_{++}, \lambda_{--}, \lambda_{+-}), 
\end{aligned}\label{eq:PTeigvals}
\end{equation}
with $\lambda_{\pm \pm}/J=-6\kappa/J \pm 2 \sqrt{(\kappa/J)^2 \pm 2i\kappa/J - 2}$, and
\begin{equation}
\begin{aligned}
S &= \begin{pmatrix}
a_{--+} & a_{-++} & a_{++-} & a_{+--} \\
-i & -i & i & i \\
ia_{--+} & ia_{-++} & -ia_{++-} & -ia_{+--} \\
1 & 1 & 1 & 1 
\end{pmatrix}, \\
	a_{\pm \pm \pm} &= 1 \pm i\kappa/J \pm i \sqrt{(\kappa/J)^2\pm 2i\kappa/J -2}\\
 &= 1+ i(3\pm1)\kappa/J + (i/2) \lambda_{\pm\pm}/J.
\end{aligned}
\end{equation}

Now, the action of the perturbation $\mathcal{V}$ on $\mathcal{Q'}$ and $\mathcal{P}$, projected respectively onto $\mathcal{P}$ and $\mathcal{Q'}$, is 
\begin{equation}
\mathcal{V}^+ = \begin{pmatrix}
-2V_{xx} & 0 & 0 & 0
\end{pmatrix}^T, \;\; \mathcal{V}^- = \begin{pmatrix}
2V_{xx} & 0 & 0 & 0
\end{pmatrix}.
\end{equation}

Putting it all together into \eqref{eq:L2eff}, we get the following final expression:
\begin{equation}
\begin{aligned}    
	-L_{2,Z_1}^{\text{eff}}/J =& \left[\frac{2J}{a_{--+} - a_{-++}} \left( \frac{a_{--+}}{\lambda_{-+}} - \frac{a_{-++}}{\lambda_{++}} \right) \right. \\  +& \left. \frac{2J}{a_{+--} - a_{++-}} \left( \frac{a_{+--}}{\lambda_{+-}} - \frac{a_{++-}}{\lambda_{--}} \right) \right]\left(V_{xx}/J\right)^2\\
    =& \frac{8(\kappa/J)^3 + 3\kappa/J}{16(\kappa/J)^4 + 9(\kappa/J)^2 + 1} \left(V_{xx}/J\right)^2.
\end{aligned}
\end{equation}%

This expression for $L_{2,Z_1}^{\text{eff}}/J$, which is equal to half the maximum spread $\delta/2$, in units of $(V_{xx}/J)^2$, is displayed  on \figref{fig:gapvskappa}[b] as a function of $\kappa/J$.

\subsection{$\mathds{Z}_2\cross \mathds{Z}_2$-symmetry-breaking perturbation}

To contrast the previous section, we look at the behaviour of the model subjected to a perturbation $H_y=V_y\sum_{l=1}^N Y_l$. This breaks the symmetry. Its action on the $Z_1$ steady-state will be similar, because it is a local state, but now, the strings of $X$ operators in $G_o$, $G_e$ and all other components of the steady state generated by them will also be affected. And it is expected for them to be coupled to many more fragments of the system and therefore the stationary subspace is expected to decay much faster than with the the symmetry-preserving perturbation. 

We take, for example, the state $G_oG_e=\bigotimes_{l} X_l$. The perturbation couples its left boundary to a 2-dimensional fragment generated by $Y_1G_oG_e$. The $Y_2$ term couples it to a 4-dimensional fragment generated by $Y_{2}G_oG_e$. The terms on sites $N$ and $N-1$ have an identical effect by symmetry. And in the bulk, each pertrubation term acting on sites from $l=3$ to $l=N-2$ couples the state to an 8-dimensional fragment generated by $Y_{l}G_oG_e$. There are $N-4$ such bulk terms. Following a similar approach as in the previous section and summing all the contributions, we obtain:
\begin{equation}
\begin{aligned}    
	&-L_2^{\text{eff}}/J = \left[ \frac{16\kappa/J}{8(\kappa/J)^2 + 1} + \frac{2}{3\kappa/J} \right. \\  +&  \left. (N-4)8\kappa/J\frac{64(\kappa/J)^2+3}{1536(\kappa/J)^4 +152(\kappa/J)^2 + 3} \right] \left(V_y/J\right)^2.
\end{aligned}\label{eq:Hyspread}
\end{equation}
We observe that the second term diverges when $\kappa/J\rightarrow0$ and that there is now a dependence on $N$ due to the bulk terms. These features can be seen on \figref{fig:gapvskappa}[b] as well. This state under the perturbation decays about one order of magnitude quicker than all the states under the symmetry-preserving perturbation, and that difference increases with system size. The other steady-states, except for the identity, $Z_{1}$, $Z_{N}$ and $Z_{1}Z_{N}$, also acquire this sort of a lifetime dependent on $N$, but the maximum shift, denoted $\delta$ on \figref{fig:spec}, is given by Eq.~\eqref{eq:Hyspread}.